\documentclass[numberedappendix]{emulateapj}
\usepackage{graphicx}
\usepackage{amssymb,amsmath} 

\newcommand\be{\begin{equation}}
\newcommand\ee{\end{equation}}

\def\msun{\,{\rm M_\odot}}
\def\gsim{ \lower .75ex \hbox{$\sim$} \llap{\raise .27ex \hbox{$>$}} }
\def\lsim{ \lower .75ex\hbox{$\sim$} \llap{\raise .27ex \hbox{$<$}} }

\begin{document}

\shorttitle{Eccentric massive black hole binaries}
\shortauthors{A. Sesana}
\slugcomment{}

\title{Self consistent model for the evolution of eccentric massive black hole binaries in stellar environments: implications for gravitational wave observations}

\author{Alberto Sesana\altaffilmark{1}}

\altaffiltext{1}{Max Planck Institute for Gravitationalphysik (Albert Einstein Institute), Am M\"uhlenberg , 14476, Golm, Germany}

\begin{abstract}
We construct evolutionary tracks for massive black hole binaries (MBHBs) 
embedded in a surrounding distribution of stars. The dynamics of the 
binary is evolved by taking into account the erosion of the central
stellar cusp bound to the massive black holes, the scattering of unbound 
stars feeding the binary loss cone, and
the emission of gravitational waves (GWs). Stellar dynamics is treated in 
a hybrid fashion by coupling the results of numerical 3-body scattering
experiments of bound and unbound stars to an analytical framework for 
the evolution of the stellar density distribution and for the efficiency
of the binary loss cone refilling. Our main focus is on the behaviour 
of the binary eccentricity, in the attempt of addressing its importance 
in the merger process and its possible impact for GW 
detection with the planned Laser Interferometer Space Antenna ({\it LISA}), 
and ongoing and forthcoming pulsar timing array (PTA) campaigns.   
We produce a family of evolutionary tracks extensively sampling the
relevant parameters of the system which are the binary mass, mass ratio and
initial eccentricity, the slope of the stellar density distribution, its
normalization and the efficiency of loss cone refilling. We find that,
in general, stellar dynamics causes a dramatic increase of the MBHB 
eccentricity, especially for initially already mildly eccentric 
and/or unequal mass binaries. This affects the overall system dynamics; 
high eccentricities enhance the efficiency of GW emission, accelerating 
the final coalescence process. When applied to standard MBHB population models, 
our results predict eccentricities in the ranges $10^{-3}-0.2$ and $0.03-0.3$ 
for sources detectable by {\it LISA} and PTA respectively. Such figures
may have a significant impact on the signal modelling, on source 
detection, and on the development of parameter estimation algorithms.    
\end{abstract}


\section{Introduction}
It is now widely recognized that massive black holes (MBH) are fundamental
building blocks in the process of galaxy formation and evolution. 
MBHs are a ubiquitous components of nearby galaxy nuclei  
\citep[see, e.g., ][]{mago98}, and their masses tightly correlate with
the properties of the host \citep[][and reference therein]{hr04}. 
In popular $\Lambda$CDM cosmologies, structure formation proceeds
in a hierarchical fashion \citep{wr78}, through a sequence of merging events. 
If MBHs are common in galaxy centers at all epochs, as implied by the
notion that galaxies harbor active nuclei for a short period of
their lifetime \citep{hr93}, then a large number 
of massive black hole binaries (MBHBs) 
are expected to form during cosmic history. The evolution
of such binaries was firstly sketched by \cite{br80}, but 
after thirty years, several details of the involved dynamical processes 
are still unclear. 

In stellar environments, the MBHB evolution proceeds via super-elastic 
scattering of surrounding stars intersecting the binary orbit \citep[slingshot
mechanism, ][]{mv92}, and the fate of the system depends on the supply of stars
available for such interaction. On the other hand, if the system is gas
rich, torques exerted by a massive circumbinary disk have been proven 
efficient in shrinking the binary down to $\sim0.1$ pc 
\citep{escala05,dotti07},
which is the current resolution limit of dedicated smoothed particle
hydrodynamical simulations. 
However, whether viscous angular momentum extraction is efficient all
the way down to coalescence is questionable \citep{lod09}.

In general, the vast majority of studies (mostly numerical) devoted
to the subject have focused on the shrinking of the binary semimajor axis,
because the relative small number of particles involved ($N<10^6$)
make the eccentricity behaviour fairly noisy. However, 
eccentricity may play an important role in the final coalescence, because
at a given semimajor axis, the coalescence timescale associated to 
gravitational wave (GW) emission is much shorter 
for very eccentric binaries \citep{pm63}. Moreover, having a 
trustworthy model for the eccentricity evolution of the system may be 
of crucial importance for the practical detection of MBHBs in the 
forthcoming GW windows.

MBHBs are infact expected to be the loudest sources of gravitational
radiation in the nHz--mHz frequency range 
\citep{Haehnelt94,Jaffe03,Wyithe03,eno04,Sesana04,
Sesana05,Jenet05,rw05,svc08,svv09}. 
The space-borne observatory 
Laser Interferometer Space Antenna \citep[{\it LISA},][]{dan08} has been
planned to cover the range of frequencies from $10^{-4}$ Hz to $0.1$ Hz.
Moving to the nanohertz frequency range, the Parkes Pulsar 
Timing Array \citep[PPTA; ][]{Manchester08}, the European Pulsar Timing Array
\citep[EPTA; ][]{Janssen08} and the North American Nanohertz Observatory
for Gravitational Waves \citep[NANOGrav; ][]{jen09} are already collecting
data and improving their sensitivity in the of $10^{-8}-10^{-6}$ Hz window, 
and in the next decade the planned Square Kilometer Array 
\citep[SKA; ][]{Lazio09} will provide a major leap in sensitivity. 

Besides the technical progresses in the instrumentation, 
the source signal modelling and the development 
 of appropriate data analysis techniques for recovering 
the sources from the data stream are crucial for the success of 
the GW astronomy challenge. So far, most of the attention was focused 
on circular MBHBs. This seems to be justified, because GW emission is 
very efficient in dumpening the binary eccentricity, and
since GW detectors ({\it LISA} in particular) are sensitive to the
very end of the MBHB inspiral, sources are expected to be circular when they 
enter the observable band. Consequently, most of the source modelling
and the signal searches and analyses, e.g. the
source injection in {\it LISA} mock data challenge \citep{bab09}, or the 
investigation carried by the {\it LISA}
parameter estimation task force \citep{aru09}, relied on
this assumption. 

However, both stellar and gas based shrinking mechanisms have 
proven to be efficient in increasing the binary eccentricity
\citep{qui96,an05,shm06,shm08,bau06,mat07,ber09,cua09,as09,as10}, 
calling into question whether the assumption of circular orbits is justified
for such GW sources. In this paper we construct a self-consistent simple
model for tracking the evolution of the MBHB eccentricity (and semimajor
axis) in {\it stellar} environments. We model the stellar distribution
surrounding the bound binary as an isothermal sphere ($\rho\propto r^{-2}$)
matching a cusp with a power law density profile $\rho\propto r^{-\gamma}$ 
inside the binary influence radius. The MBHB is evolved taking into account
the scattering of bound stars leading to the erosion of the cusp
\citep{shm08}, the subsequent scattering of unbound stars 
intersecting the binary semimajor axis \citep{qui96,shm06}, and the 
efficient GW emission stage \citep{pm63} leading to final coalescence 
of the system. The main goal of the paper is to build sensible
evolutionary tracks for the MBHB as a function of the binary mass,
mass ratio, initial eccentricity at pairing and cusp slope, and to
show that viable MBHB evolution scenarios predict a significant
eccentricity in the frequency band relevant to GW observations
with {\it LISA} and PTAs.

The paper is organized as follows. In Section 2, we extensively
describe our model, defining the relevant physical mechanisms and
writing down the evolution equations for the system. In Section 3 we
provide further insights about the physics of the model, linking our
treatment to the loss cone refilling theory. We present in detail
our evolutionary tracks in Section 4, discussing the dependencies on
the relevant model parameters, and we draw predictions for 
{\it LISA} and PTA observations in Section 5. Our main findings 
are summarized in Section 6.
 
\section{Ingredients of the model}

\subsection{Initial setup}\label{setup}
We consider a MBHB with mass $M=M_1+M_2$ ($M_1>M_2$) described by its semimajor 
axis $a$ and eccentricity $e$. 
The system is embedded in a purely stellar background
with a density profile described by a double power law, as follows:
\be
\begin{array} {l}
\rho(r)=\rho_{\rm inf}\left(\frac{r}{r_{\rm inf}}\right)^{-\gamma}\,\,\,\,\,\,r<r_{\rm inf}\\
\rho(r)=\frac{\sigma^2}{2\pi G r^2}\,\,\,\,\,\,\,\,\,\,\,\,\,\,\,\,\,\,\,\,\,\,\,\,\,\,r>r_{\rm inf}.
\end{array}
\label{rho}
\ee
Here $r_{\rm inf}$ is the influence radius of the binary, identifying the region 
where the gravitational potential is dominated by the two MBHs, and formally 
defined as the radius containing a stellar mass equal to $M$, and 
$\rho_{\rm inf}$ is the stellar density at $r_{\rm inf}$. The density
distribution is normalized to an isothermal sphere for $r>r_{\rm inf}$, such a condition
sets 
\be\label{sis}
\rho_{\rm inf}=\frac{\sigma^2}{2\pi G r_{\rm inf}^2} 
\ee
and
\be\label{rinf} 
r_{\rm inf}=(3-\gamma)\frac{GM}{\sigma^2}\approx 0.8{\rm pc}\,(3-\gamma)M_6^{1/2}, 
\ee
where $M_6$ is the total mass of the binary in units of $10^6\msun$, 
and we made use of the well established $M-\sigma$ relation in 
the form \citep{tr02}
\be\label{msigma}
M_6=0.84\sigma_{70}^4
\ee
($\sigma_{70}$ is the velocity dispersion in units of $70$km s$^{-1}$)
to get rid of $\sigma$ in the last approximation. 
We identify the region $r<r_{\rm inf}$ as the 
inner cusp, and we use $\gamma=1, 1.5, 2$
corresponding to nuclei characterized by cores/weak cusps, mild cusps 
and steep cusps respectively. 

N-body simulations of unequal mass binaries in the mass ratio range 
$0.01-10^{-3}$ \citep{bau06,mat07} have shown that dynamical friction 
is efficient in driving 
the secondary MBH much deeper than $r_{\rm inf}$ in the potential well of the 
primary-cusp system. The two MBHs pair together forming a MBHB and continue
to harden down to a separation at which the enclosed mass in the binary is
of the order of $M_2$, without
significantly affecting the stellar density profile. This is an indication 
that the hardening is still driven by the dynamical friction exerted by the 
overall distribution of stars, rather than by close individual encounters 
with stars intersecting the binary orbit. 
In our model, we assume that $M_2$ is driven by dynamical friction down 
to a separation $a_0$ where the enclosed stellar mass in the binary is 
twice the mass of the secondary MBH
\be
a_0=(3-\gamma)\frac{GM}{\sigma^2}\left(\frac{q}{1+q}\right)^{1/(3-\gamma)}=
r_{\rm inf}\left(\frac{q}{1+q}\right)^{1/(3-\gamma)},
\label{a0}
\ee
($q=M_2/M_1$ is the mass ration of the binary system) 
without affecting the stellar distribution in the cusp significantly. 
At that point, three body interactions take over, and the binary evolution
is dictated by individual encounters with stars intersecting its orbit.  

\subsection{Physical mechanisms in operation}
On its way to final coalescence starting from $a_0$, the binary is subject 
to three main dynamical mechanisms driving its evolution, namely:
(i) the erosion of the cusp bound to the primary MBH, (ii) the scattering of
unbound stars supplied into the binary loss cone by relaxation processes
once the stellar distribution is significantly modified by the MBHB, and 
(iii) the emission of GWs. The detailed description of each mechanism has 
been presented elsewhere, and the reader will be referred throughout this
section to the appropriate references for further insights. The focus
of the present work is to add them together coherently, to produce 
sensible, although admittedly very simplistic, evolutionary tracks for 
MBHBs hardening in stellar environments.

\subsubsection{Bound cusp erosion}
The hardening of an unequal mass MBHB, with an initial semimajor axis $a_0$ and
eccentricity $e_0$,
in a bound cusp was extensively studied by \cite{shm08}, hereinafter SHM08. 
Using their formalism, two differential equations 
determine the rate of change of the orbital separation and eccentricity:
\begin{equation}\label{eq:aev}
\frac{da}{dt}\big|_b=-\frac{2a^2}{GM_1M_2}\int_0^\infty \Delta {{\cal E}}
\frac{d^2N_{\rm ej}}{da_* dt}da_*,
\end{equation}
\begin{equation}\label{eq:eev}
\frac{de}{dt}\big|_b=\int_0^\infty \Delta {e}\frac{d^2N_{\rm ej}}{da_* dt}da_*. 
\end{equation}
Here $a_*$ is the semimajor axis of a star bound to $M_1$ and 
$d^2N_{\rm ej}/da_* dt$ is the number of stars subject to slingshot ejection 
in the semimajor axis and time intervals $[a_*, a_*+da_*]$, $[t, t+dt]$. 
The terms $\Delta e$, $\Delta {\cal E}$ are measured from scattering 
experiments. The ejection rate $d^2N_{\rm ej}/da_* dt$ is instead computed by
coupling the numerical results of the experiments to an analytic framework
for the binary evolution, which is embedded in a cusp of the form described
by equation (\ref{rho}), as detailed in SHM08. 
The major finding of SHM08 is that the binary hardens by a factor of $\sim10$  
by extracting the binding energy of the stars in the cusp. During this 
process, $e$ usually increases by a large factor, depending on the binary
mass ratio and on the cusp slope. Results are tabulated in table 1
of SHM08.

 
\subsubsection{Slingshot of unbound stars}
The theory of MBHB hardening in a distribution of unbound field stars 
characterized by a density $\rho$ and a velocity dispersion $\sigma$  
was singled out  by \cite{qui96} and extensively revisited by 
\cite{shm06}, hereinafter SHM06.
The binary evolution can be expressed as a function of the
dimensionless hardening rate $H$ and eccentricity growth rate $K$ as
\begin{equation}\label{eq:aev}
\frac{da}{dt}\big|_u=-\frac{a^2G\rho}{\sigma}H,
\end{equation}
\begin{equation}\label{eq:eev}
\frac{de}{dt}\big|_u=\frac{aG\rho}{\sigma}HK. 
\end{equation}
The quantities $H$ and $K$ are related to the average energy and 
angular momentum exchange between the stars and the binary in a single
encounter, and are
computed via extensive three body scattering experiments, as described,
e.g., in SHM06. In general, hardening by scattering of unbound stars
becomes effective when the binary reach the so called hardening 
radius, defined as \citep{qui96}
\be\label{ah}
a_h\approx\frac{GM_2}{4\sigma^2}.
\ee 
This is the separation at which the specific binding energy of 
the binary is of the order of the specific kinetic energy of the field
stars. For $a>a_h$, stars are basically too fast to effectively exchange
energy and angular momentum with the binary (soft binary
regime); when $a<a_h$, the binary tends to capture stars in short living 
weakly bound orbits, kicking them to infinity with $v>\sigma$ (hard binary
regime). The transition soft/hard binary is rather smooth, and happens
at about $a_h$. Once the binary is hard, its hardening proceeds at about 
constant rate, as shown by the $H$ tracks plotted in figure 3 of SHM06. 
Perfectly circular binaries tend to stay circular (because of the 
conservation of the Jacobian integral of motion in the 3-body problem), 
while even slightly eccentric binaries tend to increase their eccentricity, 
as shown by the $K$ rates plotted in figure 4 of SHM06.
 
\subsubsection{Gravitational wave emission}
For our purposes, the effect of GW emission can be modelled in the quadrupole 
approximation. Under this assumption, the evolution equations
for the system are given by \cite{pm63}
\begin{eqnarray}\label{dadtgw}
\frac{da}{dt}\big|_{\rm gw}&=&-\frac{64}{5}\frac{G^3}{c^5}\frac{M_1M_2M}{a^3(1-e^2)^{7/2}}\left(1+\frac{73}{24}e^2+\frac{37}{96}e^4\right)\nonumber\\
&=&-\frac{64}{5}\frac{G^3}{c^5}\frac{M_1M_2M}{a^3}F(e)
\end{eqnarray}
\be
\frac{de}{dt}\big|_{\rm gw}=-\frac{304}{15}\frac{G^3}{c^5}\frac{M_1M_2M}{a^4(1-e^2)^{5/2}}e\left(1+\frac{121}{304}e^2\right).
\ee
The function $F(e)$ is defined by the last equality in equation \ref{dadtgw}.
The shrinking rate is a strong factor of $a$, meaning that GW-driven hardening
is effective only at small separations. The eccentricity evolution rate is 
also a strong function of $a$ and $e$ itself, and it is always negative. 
GW emission, therefore, is very effective in circularizing MBHBs, which, in
turn, is the reason why little attention has been paid so far to eccentric
systems in the context of GW detection.

\subsection{General equations for the binary evolution}
Having identified the relevant mechanisms at play, we can put the 
pieces together by writing the evolution of the binary as
\begin{equation}\label{aevt}
\frac{da}{dt}=\sum_i\frac{da}{dt}\big|_i
\end{equation}
\begin{equation}\label{eevt}
\frac{de}{dt}=\sum_i\frac{de}{dt}\big|_i,
\end{equation}
where $i=b,u,{\rm gw}$ labels the three mechanisms considered. Here we
handle MBHBs in stellar environments, and the relevant scale of the 
stellar distribution is defined by the sphere of influence $r_{\rm inf}$
of the massive black hole binaries. It is then natural to consider 
as relevant parameters the stellar density and velocity dispersion at
the influence radius, $\rho_{\rm inf}$ and $\sigma_{\rm inf}=\sigma$ 
(for an isothermal
sphere the velocity dispersion is independent on radius). It is then 
instructive to recast the MBHB dynamics in the dimensionless $H$ and $K$ 
formalism proposed by Quinlan, to compare the dimensionless rates 
given by each mechanism.
The global evolution of the system can be then written as a generalization
of equations (\ref{eq:aev}) and (\ref{eq:eev}) in the form 
\be\label{htot}
\frac{da}{dt}=-\frac{a^2G\rho_{\rm inf}}{\sigma}\sum_iH_i
\ee
\be\label{ktot}
\frac{de}{dt}=\frac{aG\rho_{\rm inf}}{\sigma}\sum_iH_iK_i,
\ee
where $H_i$ is trivially defined as 
\be
H_i=\frac{\sigma}{a^2G\rho_{\rm inf}}\frac{da}{dt}\big|_i
\ee
and 
\be
K_i=a\frac{de}{dt}\big|_i\left(\sum_j\frac{da}{dt}\big|_j\right)^{-1}.
\ee
We integrate the coupled differential equations (\ref{htot}) and 
(\ref{ktot}) starting from $a_0$. As described in Section \ref{setup}
(equation (\ref{a0})), the 
value of $a_0$ is set by the total mass of the binary $M$, the mass
ratio $q=M_2/M_1$, the cusp slope $\gamma$ and the stellar
velocity dispersion $\sigma$. In our default models
we force $\sigma$ to obey the $M-\sigma$ relation given by equation 
(\ref{msigma}). In  this manner there is a one-to-one correspondence 
between the mass of the binary and $\sigma$, i.e., equally massive 
binaries are embedded in identical isothermal spheres. 
Having set the normalization of the stellar distribution and the
initial separation $a_0$, the evolution of the binary depends on the
four parameters $M_1$, $q$, $e_0$ and $\gamma$. We extensively
sample this parameter space as following:
\begin{itemize}
\item ${\rm log}(M_1/{\rm M}_{\sun})=2,3,4,5,6,7,8,9,10,11$
\item $q=1, 1/3, 1/9,...,1/729$
\item $e_0=0.01,0.1,0.3,0.6,0.9$
\item $\gamma=1, 1.5, 2$
\end{itemize}
for a grand total of  $10\times7\times5\times3=1050$ simulations.
Even though the binary evolution under the effect of stellar encounters 
is basically scale free, the simulation of systems with different 
absolute masses was necessary to match together the scattering-driven 
phase to the GW-driven phase, which instead is highly mass dependent.
The assumption of different eccentricities at the moment of pairing
takes the environmental effects affecting the dynamical
friction stage into account. In general, during the merging process, 
galaxies capture
each other on a very eccentric orbit, which is reflected in the initial
trajectories of the two MBHs (still at kpc separations at this stage).
Dynamical friction against massive, rotationally supported, 
circumbinary disks has been proven to circularize the orbit 
\citep{dotti06}. However, this is not in general true 
in gas poor environments, where the 
process is driven by interaction with the stellar distribution
\citep{cmg99},
and the eccentricity of the MBHB at the moment of pairing may retain
memory of its initial value, or may, in general, be different than
zero. 

We also consider four alternative models to address the impact 
of the assumed $M-\sigma$ relation and the choice of normalizing 
the efficiency of unbound scatterings to $\rho_{\rm inf}$. Let us 
denote with $\hat{\sigma}$ the value of the velocity dispersion
predicted by the $M-\sigma$ relation for a given $M$.
We consider models with velocity dispersions equal to $0.7\hat{\sigma}$
and $1.3\hat{\sigma}$, which is approximately the range of variance of 
$\sigma$ for a given MBH mass measured in the $M-\sigma$ relation
\citep{hr04}. We also run models
with two different normalizations for the unbound scattering process: a fast
model normalized to $10\rho_{\rm inf}$ and a slow model normalized to 
$0.1\rho_{\rm inf}$. The motivation for this set of runs will be clarified
in Section \ref{lcref}.

To practically evolve the binary we make use of the results of the 
scattering experiments with unbound and bound stars performed in 
SHM06 and SHM08. In those papers, the quantities $\Delta e$, $\Delta {\cal E}$, 
$d^2N_{\rm ej}/da_* dt$ (for the bound scatterings), $H$ and $K$ (for the
unbound scatterings) were recorded on a grid of $a$ and $e$, covering
the relevant dynamical range. The evolution of the coupled differential
equations (\ref{htot}) and 
(\ref{ktot}) is performed by interpolation over the grid as the binary evolves.
In SHM06, unbound scatterings were carried out for all the
considered mass ratios down to $q=1/243$. To complete the sample we 
carried out additional experiments for the case $q=1/729$. SHM08, instead,
focused on unequal MBHBs, with $q\lsim0.1$. To complete the mass
ratio sample, we ran experiments for the cases $q=1, 1/3$. We then 
have all the bound and unbound scattering experiments results spanning 
the $q$ and $e_0$ range of interest. 

\subsection{Limitations and caveats}\label{caveats}

Our evolutionary tracks are computed in a self-consistent way, summing 
together the effects of different mechanisms. However, we should be aware
of the several limitations and simplifications we have adopted. 
One major caveat is the extension of the bound scattering experiments to
mass ratios of $q>1/9$. It is infact unlikely that, in such cases, $M_2$
would reach $a_0$ without affecting the stellar cusp at all.
Cusp disruption would start earlier (especially in the equal mass 
case), and the cusp erosion phase described here may not be a trustworthy
description of reality. However, we find that the
impact of the bound cusp erosion on the binary evolution is smaller for
larger mass ratios. This is because for $q\rightarrow1$, $a_0\sim r_{\rm inf}$, 
(see figure \ref{fscale}) and the impact of the binding energy extraction 
in the total energy budget of the system becomes less significant. 
The eccentricity evolution in the cusp erosion phase is only mild when 
$q=1,1/3$, implying that our approximate treatment would not 
significantly affect the overall results. 
Another caveat to bear in mind is that 
the three body scattering is a scale free problem as long as 
$m_*\ll M_2$. Even though we compute 'a posteriori' evolutionary tracks
for systems with $M_1=100\msun$ and $q=1/729$, we consider our results
meaningful only when $M_2>100\msun$. This is also reasonable, since 
we are interested in the evolution of MBHBs. Moreover, if $M_2$ is small
($<10^4\msun$), the amount of stars interacting with the binary is  
also quite small, i.e. the granularity of the problem increases. Our 
smooth evolutionary tracks should then be interpreted more as 'trends', 
or 'mean evolutions' rather than paths followed by each individual binary. 
We have also not included the possibility of stellar tidal disruption, which 
has been shown to be an efficient process in the cusp erosion phase, 
especially in the mass ratio range $0.01<q<0.1$ \citep{chen09}. 
In general, the inclusion of tidal disruptions mildly enhance the eccentricity
increase \citep{chen10}, because disrupted stars preferentially
have $a_*<a$, i.e. they would drive binary toward circularization if
ejected (see SHM08 for a detailed discussion of this effect). 
Lastly, there is a somewhat net distinction between 
bound and unbound scatterings in our formalism, 
which is certainly oversimplistic, since in 
reality relaxation processes will mix-up the different stellar populations. 
The appearance of distinctive features in the transition between the bound 
and the unbound regime can be therefore considered somewhat artificial, 
the reality would probably be more gentle. We do not believe that this 
has a major impact on our main results.

\section{Physics of the binary evolution}
\begin{figure}
\centering
\includegraphics[width=0.95\linewidth]{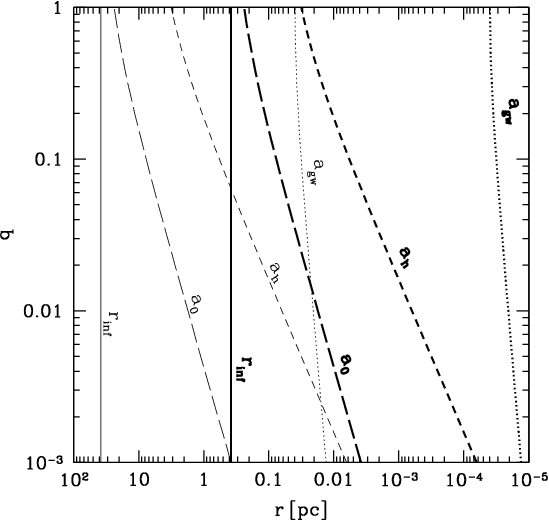}
\caption{Relevant lengthscales in the binary hardening problem as a 
function of the binary mass ratio for two different absolute values
of the binary mass $M=10^5\msun$ (thick lines) and $M=10^9\msun$ 
(thin lines). For each set of curves, from left to right, we plot
$r_{\rm inf}, a_0, a_h$ and $a_{\rm gw}$ (for a circular binary, i.e.
assuming $F(e)=1$, see equation (\ref{dadtgw})), 
as labelled in the figure. The cusp slope is fixed to $\gamma=1.5$.}
\label{fscale}
\end{figure}

The relative contributions of the three mechanisms considered in the 
previous section are set by the typical lengthscales below which they 
become effective. 
As stated before, cusp erosion becomes effective at $a_0$. At this 
point $H_b$ and $K_b$ start to dominate the MBHB evolution. 

On the other hand, scattering of unbound stars becomes fully effective 
at the hardening radius $a_h$ defined by equation (\ref{ah})
\begin{eqnarray}\label{ah}
a_h & \approx & \frac{1}{4(3-\gamma)}
\left(\frac{q}{1+q}\right)^{(2-\gamma)/(3-\gamma)}a_0\nonumber\\
& \approx & 0.2{\rm pc}\,M_6^{1/2}\left(\frac{q}{1+q}\right).
\end{eqnarray}
We see that $a_h=(1/4)a_0$ independently of the mass ratio for 
$\gamma=2$, and in general,
with decreasing $\gamma$, the ratio $a_h/a_0$ becomes smaller and $q$ 
dependent. As the binary shrinks, the central cusp is depleted and the
scattering of unbound stars refilling the binary loss cone (i.e., the family
of stellar orbits intersecting the binary semimajor axis, see
next section) becomes dominant.
In this second stage, occurring at approximately $0.1a_0$ the binary
evolution is determined by $H_u$ and $K_u$. 

Gravitational radiation will eventually take over at $a_{\rm gw}$, driving the
binary to the final coalescence. $a_{\rm gw}$ can be derived
by imposing $da/dt|_u=da/dt|_{\rm gw}$ ; rearranging and dividing
by $a_h$ we find
\begin{eqnarray}\label{agw}
a_{\rm gw}&=&4\left[\frac{128\pi(3-\gamma)F(e)}{5H}\right]^{1/5}\frac{\sigma}{c}q^{-4/5}(1+q)^{3/5}a_h\nonumber\\
&\approx&0.00027{\rm pc}\,(3-\gamma)^{1/5}F(e)^{1/5}M_6^{3/4}\frac{q^{1/5}}{(1+q)^{2/5}},
\end{eqnarray}
where $F(e)$ is defined by equation (\ref{dadtgw}),
and we again used the $M-\sigma$ relation to write the last approximation.
Figure \ref{fscale} highlights the behaviour of the lengthscales of the
system ($r_{\rm inf}, a_0, a_h, a_{\rm gw}$) as a function of $q$ for two
selected values of $M$ and $\gamma=1.5$. 
It is easy to see that, in general, $a_{\rm gw}\ll a_h$. 
For example, assuming
circular binaries with $q=1$, $H=15$, and $\sigma=100$km/s 
(or $M=4\times 10^6\msun$, according to the $M-\sigma$) gives 
$a_{\rm gw}/a_h\approx 2\times 10^{-3}$. $a_{\rm gw}$ can approach $a_h$ if 
the mass ratio is extreme ($q\sim 10^{-3}$) and the velocity dispersion is 
very large (e.g. if the binary is massive; $\sigma>300$km/s or $M>10^9\msun$),
as shown by the set of thin lines in figure \ref{fscale}. 
In general, figure \ref{fscale} clearly shows that there are three well 
defined zones where one single shrinking mechanism is dominant among 
the others, and we have:
\be
a_{\rm gw}<a_h<a_0<r_{\rm inf}.
\ee

\subsection{$H$ and $K$ formalism in the loss cone framework}
It is instructive at this point to frame the hardening rate given by equation 
(\ref{eq:aev}) in the context of loss cone refilling theory 
\citep{fr76,ls77,ck78}.
After the binary has depleted all the stars intersecting its orbit
(formally defining what is called the 'binary loss cone' of the stellar 
distribution function, see \cite{mm03} for a comprehensive review), 
its further hardening depends on the rate $\Gamma$ 
at which such orbits (i.e the loss cone) are refilled. Loss cone refilling 
can proceed by diffusion of stars either in energy ($\epsilon_*$) or in 
angular momentum ($j_*$). In general, stellar encounters are much 
more efficient in changing the star angular momentum, and diffusion in 
the $j_*$ space is the relevant process. The loss cone refilling depends 
on the average $\Delta{j_*}$ experienced by a star on a almost radial orbit
during one orbital period. If this change is larger than the size of the 
binary loss cone in the angular momentum space, $j_{\rm lc}\sim\sqrt{2GMa}$, 
then stars are easily scattered back and forth into the loss cone, and the 
loss cone is filled. In this regime, the supply rate of stars from a given 
distance to the binary $r$ is given by \citep{ls77,pa08}

\be\label{dgamma}
\frac{d\Gamma_f}{d{\rm log}r}\approx\frac{a}{r}\frac{N_*(<r)}{P(r)}.
\ee
Here $P(r)$ is the typical period of a star on an almost radial 
orbit coming from a distance $r$ and $N_*(<r)$ is the number of stars enclosed
in a sphere of radius $r$ around the MBHB. Let us focus on stars coming from
$r>r_{\rm inf}$. Assuming an isothermal sphere $N_*(<r)=(M/m_*)(r/r_{\rm inf})$,
and that the typical period of a star on a radial orbit is $P(r)\sim r/\sigma$, 
integrating equation (\ref{dgamma}) from $r_{\rm inf}$ to $\infty$, and
using equation (\ref{sis}), we get
\be\label{gammaf}
\Gamma_f\approx 2\pi\frac{M}{m_*}\frac{G\rho_{\rm inf}a}{\sigma}.
\ee

Let us contrast this result with equation (\ref{eq:aev}).
In the Quinlan formulation, the binary is embedded in a homogeneous stellar
field with density $\rho$. The hardening rate $H$ is then derived writing 
the interaction rate as a flux of stars through the binary cross section, 
namely
\be
\Gamma_Q=\frac{\rho}{m_*}\Sigma v,
\ee 
where $\rho/m_*$ is the number density of field stars, $v$ their velocity
at infinity (i.e., far from the binary) and $\Sigma$ the binary cross section. 
If $b$ is the star impact parameter at infinity, and 
if we assume the encounter to be relevant only for $b<b_{\rm max}$, 
then $\Sigma=\pi b^2_{\rm max}$. Relating $b$ to the maximum approach $x$ to 
the binary via gravitational focusing ($b^2=2GMx/v^2$), and replacing
$x_{\rm max}=a$ (stars have to cross the binary semimajor axis to 
exchange energy and angular momentum efficiently), we get
\be\label{gammaq}
\Gamma_Q=2\pi\frac{M}{m_*}\frac{G\rho a}{v}.
\ee 
By comparing equation (\ref{gammaf}) and (\ref{gammaq}), if we identify
the intruder velocity at infinity $v$ with the dispersion velocity in the
isothermal sphere $\sigma$, we see that our $H_u$ and $K_u$ prescriptions for 
the scattering of unbound stars correspond to considering the loss cone 
always full at $r_{\rm inf}$. 

\subsection{Loss cone refilling}\label{lcref}
By normalizing $H_u$ and $K_u$ to $\sigma$ and $\rho_{\rm inf}$, our model
implicitly assumes that the loss cone is always full at $r>r_{\rm inf}$ (i.e.,
in the so called {\it pinhole} regime), and empty for $r<r_{\rm inf}$ 
(i.e., in the so called {\it diffusive} regime, \citealt{ck78}). 
The status of the loss cone
then enter as a {\it parameter} in our formulation, set by our choice 
of normalizing the system to $\rho_{\rm inf}$. The issue of what is the
physical mechanism that keeps the loss cone full at $r>r_{\rm inf}$ is not 
addressed in this paper. We will only briefly discuss here the 
plausibility of such
scenario. After the loss cone is depleted, in absence of any other 
physical mechanism, two body relaxation \citep{bt87} sets the 
timescale for loss cone refilling. This is usually longer than
the Hubble times in real galaxies \citep{ms06}, and under
this assumption, the loss cone is, in general, in the diffusive regime way
beyond $r_{\rm inf}$. However, in more realistic situations, a myriad of 
other physical factors play a substantial role, shortening the loss cone
refilling timescale. It has been shown that axisymmetry and in particular
triaxiality \citep{yu02,mp04,ber06} are very effective in repopulating 
the loss cone, driving
stars on chaotic and centrophilic orbits toward the black hole. Moreover, 
the matter distribution in stellar bulges is far from being smooth.
Inhomogeneous concentrations of matter, such as massive star 
clusters or giant molecular clouds (the so called massive perturbers), have 
been proven efficient in perturbing stellar orbits, significantly shortening 
the loss cone refilling timescale \citep{pa08}.
We should note that these mechanisms are likely to operate efficiently for 
$r>r_{\rm inf}$. Inside the MBH influence radius, indeed, the presence of 
a central massive object dominating the gravitational potential tends to 
force the stellar system toward a more spherical symmetry (and the 
triaxial structure may be erased in the central region, see e.g \cite{mq98}). 
For the same reason, massive perturbers 
tend to be stripped by the MBH tidal field, and would hardly survive 
in this region. Also the two body relaxation timescale increases for 
decreasing $r$ if the potential is dominated by a central object, because
the velocity dispersion of the system, which in this region is the Keplerian 
velocity around the MBH, scales as $r^{-1/2}$, and the relaxation timescale
has a strong dependence on the velocity dispersion \citep{bt87}.
Moreover, inside $r_{\rm inf}$, the ejection of stars bound in the cusp 
depletes those orbits with energy $\epsilon_*<GM/(2r_{\rm inf})$, 
and since $\epsilon_*$ diffusion is much less efficient than $j_*$ 
diffusion, the contribution to the loss cone refilling 
coming from $r<r_{\rm inf}$ should be, in general, negligible. 
It is then reasonable to assume a full loss cone for $r>r_{\rm inf}$, 
and otherwise empty. We shall test the implication of such hypothesis 
by varying the normalization used in equations (\ref{htot}) and (\ref{ktot}). 
We therefore test models normalized to $10\rho_{\rm inf}$ and 
$0.1\rho_{\rm inf}$, which correspond to consider the loss cone full only 
for $r>10^{1/2}r_{\rm inf}$, or down to $r=10^{-(3-\gamma)/2}r_{\rm inf}$ (where
$\gamma$ is the slope of the inner cusp as defined by equation (\ref{rho}))
respectively. The consideration of smaller values of the relevant
density assumed for the diffusion process ($0.1\rho_{\rm inf}$) serves also
as a test for the robustness of our results against different outer density
profiles. Although our refilling rate is derived for an isothermal 
distribution, that reasonably fit the measured density profiles of 
 severa nearby galaxies and of the Milky Way \citep{l95,db98}, 
many other galaxies show shallower outer density profiles 
\citep{f94,g96,r01}
and the bulk of stars participating to the refilling mechanism may come 
from slightly larger radii, where the density is lower. We will show 
(section 5.2 and figure 7) that this have a minor impact on our results,
changing the eccentricity of the system in the observable GW bands
by a factor $\lsim 2$.

\section{Evolutionary tracks for MBHBs}

\begin{figure*}
\centering
\begin{tabular}{cc}
\includegraphics[width=0.46\linewidth]{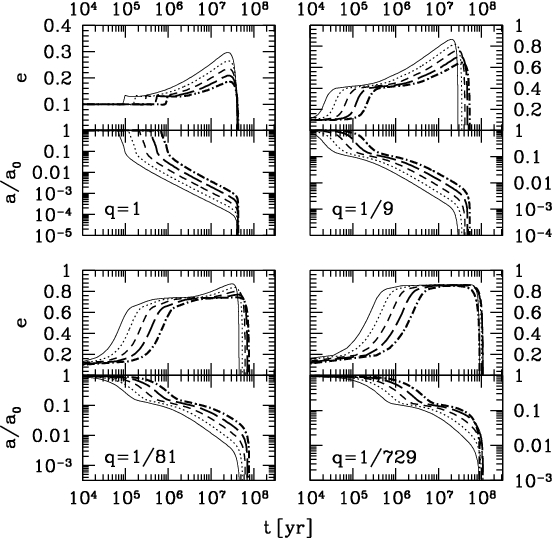} &
\includegraphics[width=0.46\linewidth]{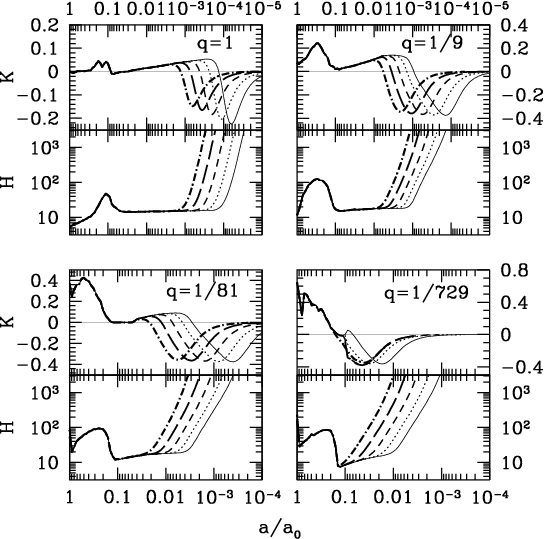} \\
\end{tabular}
\caption{MBHB evolutionary tracks produced by our model by assuming $\gamma=1.5$
and $e_0=0.1$. {\it Left plot}:
in each of the four pairs of panels we plot the eccentricity (top) and 
semimajor axis (bottom) evolution as a function of time. Different panels refer
to different values of $q$, as marked by the inset labels. Different linestyles
correspond to different binary masses: $M=10^5\msun$ (solid), 
$10^6\msun$ (dotted), $10^7\msun$ (short--dashed), $10^8\msun$ (long--dashed), 
$10^9\msun$ (dotted--dashed). {\it Right plot}: corresponding binary 
eccentricity growth rate $K$ (top panels in each of the four sectors) 
and hardening rate $H$ (bottom panels in each of the four sectors), 
as a factor of the binary separation normalized to $a_0$. 
Linestyle as in the left plot.}  
\label{figm}
\end{figure*}
\begin{figure*}
\centering
\begin{tabular}{cc}
\includegraphics[width=0.46\linewidth]{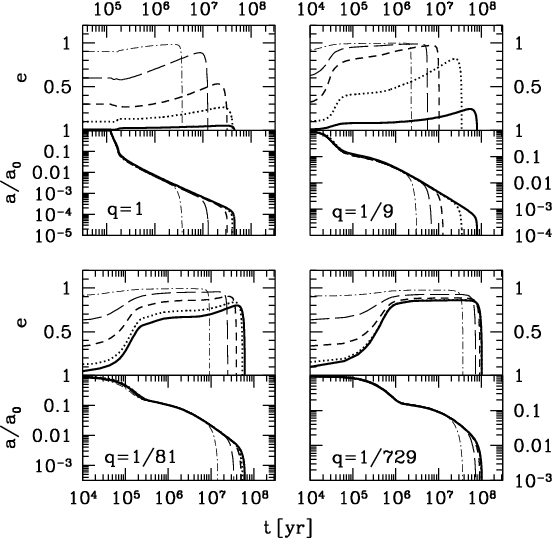} &
\includegraphics[width=0.46\linewidth]{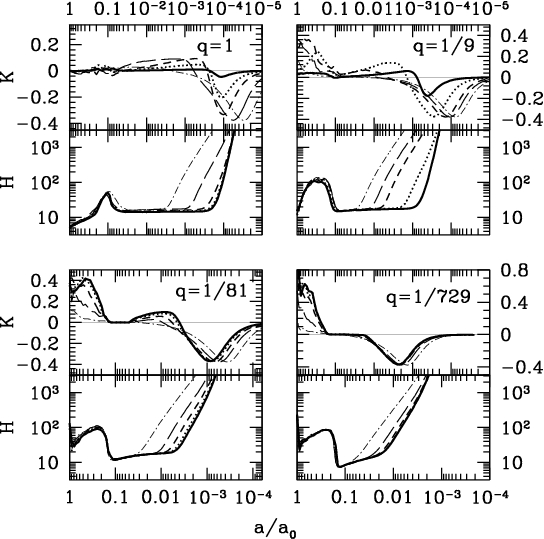} \\
\end{tabular}
\caption{Same as figure \ref{figm}, but now assuming $M=10^6\msun$
and varying the initial eccentricity $e_0$. In each individual panel
different linestyles are for $e_0=0.01$ (solid), $0.1$ (dotted), 
$0.3$ (short--dashed), $0.6$ (long--dashed), $0.9$ (dotted--dashed).
The inner cusp slope is fixed to $\gamma=1.5$.}  
\label{figecc}
\end{figure*}
\begin{figure*}
\centering
\begin{tabular}{cc}
\includegraphics[width=0.46\linewidth]{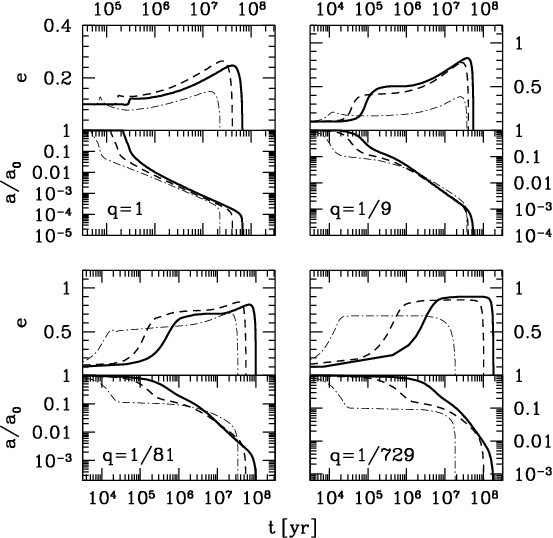} &
\includegraphics[width=0.46\linewidth]{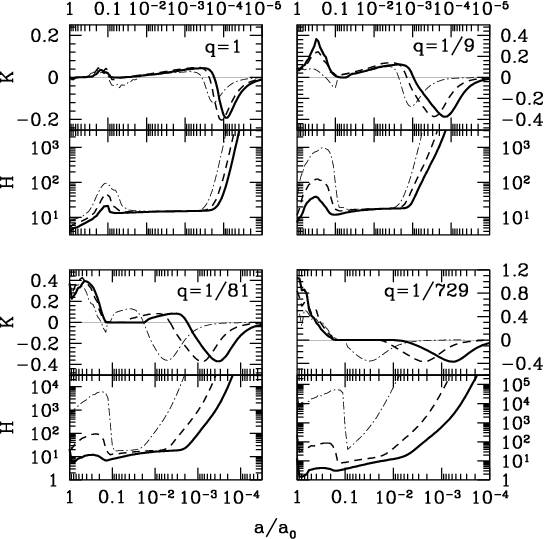} \\
\end{tabular}
\caption{Same as figure \ref{figm} but now assuming $M=10^6\msun$,
$e_0=0.1$ and varying the cusp slope. Different linestyles are for
$\gamma=1$ (solid), 1.5 (dashed), 2 (dotted--dashed).}  
\label{figgamma}
\end{figure*}

In this section we present the MBHB evolutionary tracks produced by our
hybrid model. As discussed in the previous section, the binary goes 
through three subsequent phases, which are in general distinct. In the
discussion we will simply identify them as the bound phase (erosion of the
bound cusp), unbound phase (scattering of unbound stars refilling the loss
cone), and GW phase (where GW emission become more efficient then the unbound
scattering). Each phase is characterized by its proper $H_i$ and $K_i$.
In the following compilation of plots we will present the evolution of 
the global rates $H=\sum_i H_i$ and $K=\sum_i K_i$, it will be clear by 
looking at the figures which particular mechanism dominates in each region
of the binary evolution. Each of the figures \ref{figm}, \ref{figecc} and 
\ref{figgamma} shows the quantities $e(t)$, $a/a_0(t)$, $K(a/a_0)$ and
$H(a/a_0)$ for different $q$, as labeled in each panel. In the discussion
we will simply refer to the panels as $e$, $a$, $H$ and $K$ panels.

\begin{figure*}
\centering
\begin{tabular}{cc}
\includegraphics[width=0.45\linewidth]{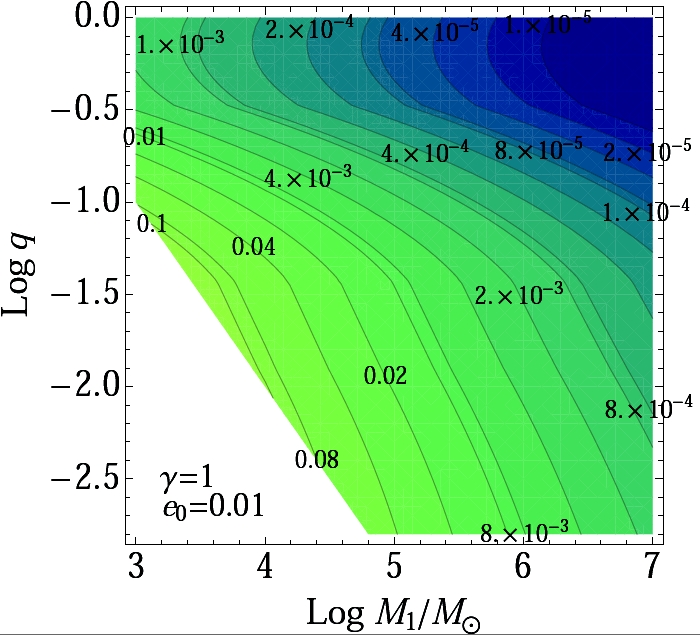} &
\includegraphics[width=0.45\linewidth]{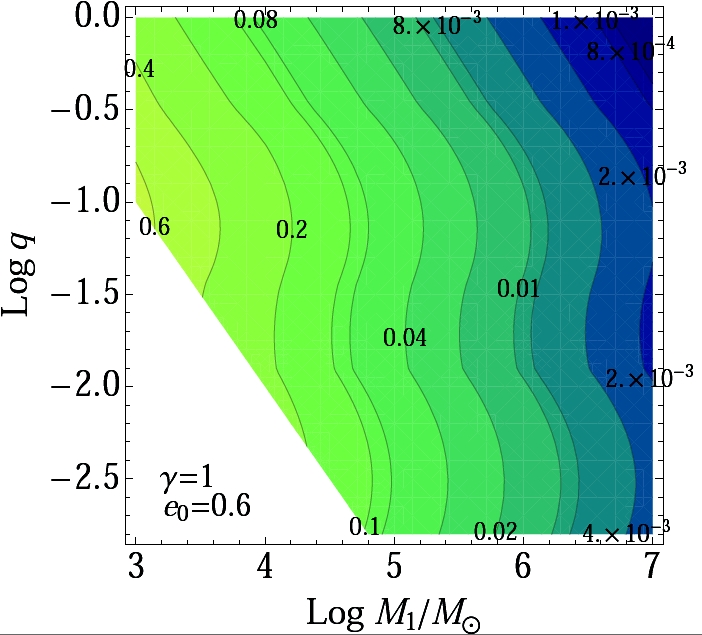} \\
\includegraphics[width=0.45\linewidth]{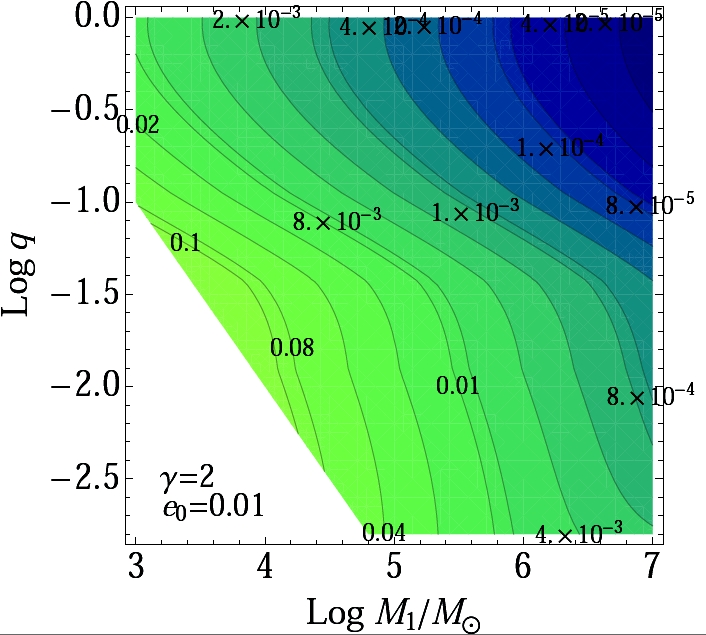} &
\includegraphics[width=0.45\linewidth]{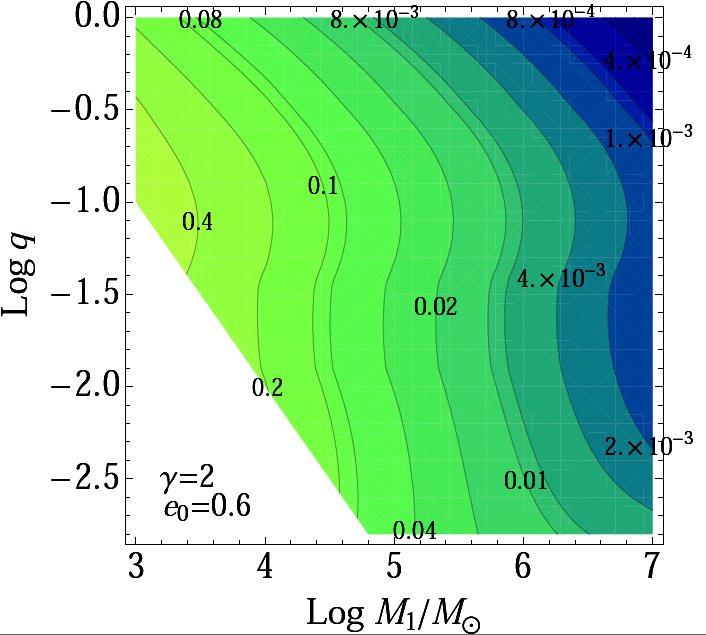} \\
\end{tabular}
\caption{Contour plots of $e$ computed at $f_{LISA}=5\times10^{-5}$Hz, 
in the $(M_1,q)$ plane
for selected model parameters, as labelled in each panel. Note that 
we excluded the bottom-left region, corresponding to systems for 
which $M_2<100\msun$.}  
        \label{figlisa}
\end{figure*}
\begin{figure*}
\centering
\begin{tabular}{cc}
\includegraphics[width=0.45\linewidth]{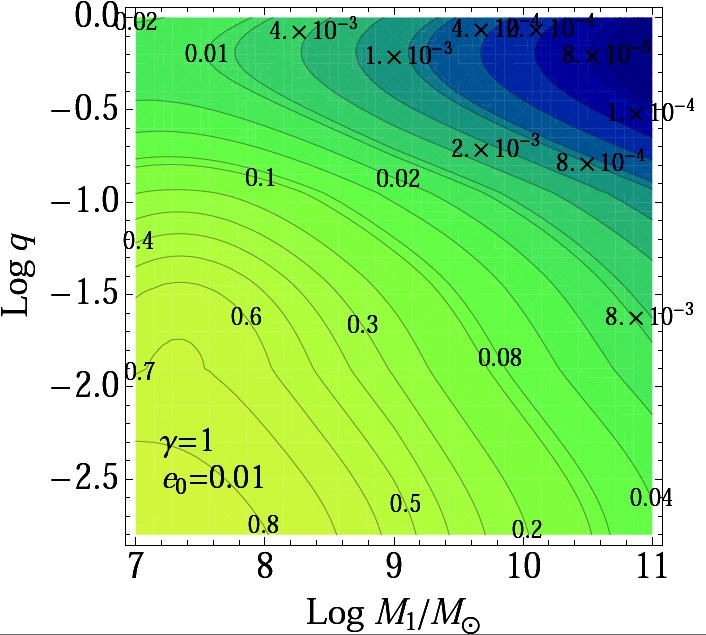} &
\includegraphics[width=0.45\linewidth]{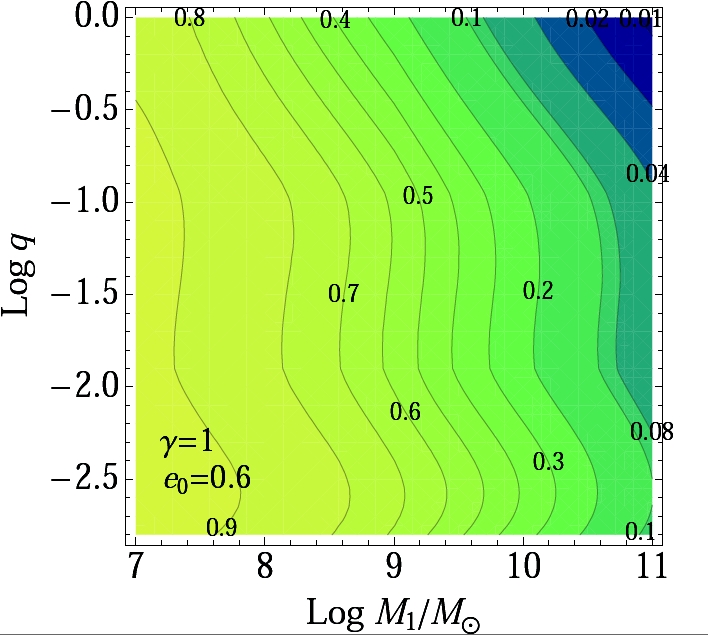} \\
\includegraphics[width=0.45\linewidth]{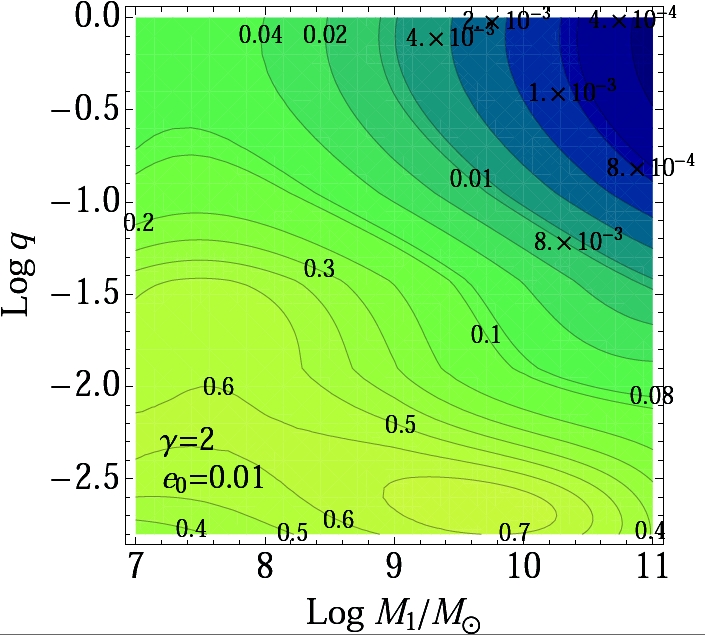} &
\includegraphics[width=0.45\linewidth]{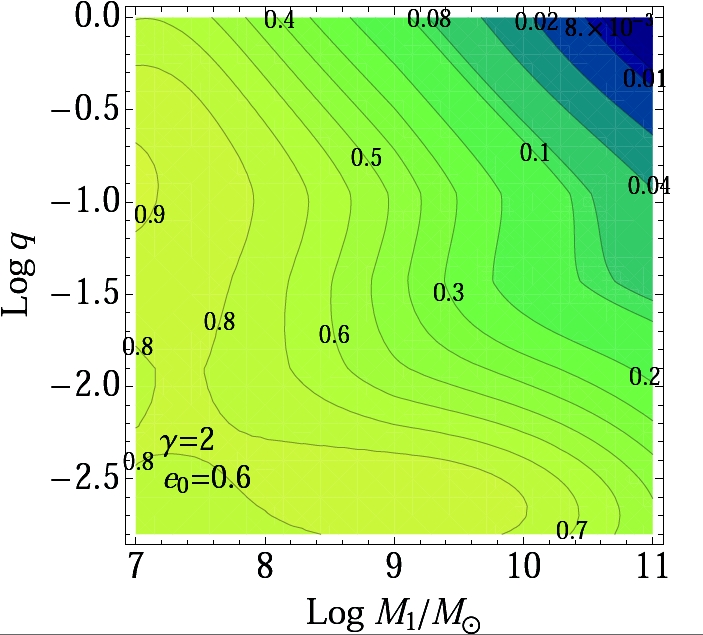} \\
\end{tabular}
\caption{Same as figure \ref{figlisa}, but now for 
$f_{\rm PTA}=5\times10^{-9}$Hz.
Note the different scale in the $x$-axis, dictated by the fact that PTAs
are sensitive to more massive systems emitting at lower frequencies.}  
        \label{figpta}
\end{figure*}

\subsection{Dependence on the model parameters}
The evolution of the system depends on the chosen values for the parameters
$M_1$, $q$, $e_0$ and $\gamma$. Such dependencies are extensively illustrated
in figures \ref{figm}, \ref{figecc} and \ref{figgamma}. Let us consider each
parameter separately, by starting with those defining the masses
of the system: $M_1$ and $q$. 

The evolution of the MBHB as a function of
$M_1$ and for different $q$ is plotted in figure \ref{figm}, assuming
$\gamma=1.5$ and $e_i=0.1$. Since our treatment of stellar 
scattering is scale free, the value of $M_1$  
affects the system evolution {\it only}, by setting the relative gap 
between $a_h$ and $a_{\rm gw}$, which has a mild dependence on $M_1$. 
By substituting the $M-\sigma$ relation in equation (\ref{agw}) we have infact 
$a_h/a_{\rm gw}\propto M^{-1/4}$, i.e. the gap is larger for lighter systems.
This means that, in general, lighter binaries become more eccentric, because,
after the bound scattering phase (which is basically scale free by construction)
they evolve under the effect of unbound scattering for a larger portion of 
their dynamical range, as it becomes clear by looking at the $K$ panels
of figure \ref{figm}. For equal mass binaries with $e_0=0.1$, the eccentricity
grows only to $0.15$ for $M_1=10^9\msun$, and up to $0.3$ for $M_1=10^5\msun$.
The mass dependence of the eccentricity growth is also evident for binaries 
with $q=1/9$, while it tends to disappear for lower mass ratios where 
the eccentricity evolution is dominated by the bound phase.
The $q$ dependence of the eccentricity evolution is emphasized by the four 
different quadrants of figures \ref{figm}, \ref{figecc} and \ref{figgamma}.
Let us consider again figure \ref{figm}. The main result here is that 
the eccentricity growth in the bound phase, at least when $e_0$ is small, 
is in general much larger for lower values of $q$, as explained in detail
in Section 4.1 of SHM08. This is nicely shown by the $K$ panels: as
$q$ decreases from 1 to $1/729$, the peak in the $K$ rate increases from
$\sim0.05$ to $\sim0.6$. The value of $q$ also sets the relative weight 
of the bound and of the unbound phases in the hardening process. Although      
$a_0/a_h$ is only mildly dependent on $q$ (depending on the cusp slope $\gamma$,
see equation (\ref{ah})), $a_h/a_{\rm gw}$ is a strong function of $q$,
(see equation (\ref{agw})) and it can be even less then one for high $M_1$ and
small $q$ (as shown in figure \ref{fscale}). Therefore, as $q$ decreases,
the eccentricity growth is dominated by the bound phase. For $e_0=0.1$,
the maximum eccentricity reached by the binary at the end of the stellar
driven phase increases from $\sim0.2$ for $q=1$ to $\sim0.9$ for $q=1/729$. 
The trends with $M_1$ and $q$ presented in figure \ref{figm} are preserved
when changing $\gamma$ and $e_0$.

The dependence of the MBHB evolution on $e_0$ is studied in figure \ref{figecc},
where we fixed $M_1=10^6\msun$ and $\gamma=1.5$. The binary eccentricity, in
general, tends to increase in the scattering phase regardless on the value
of $e_0$. When $e_0=0.01$, binaries with $q>0.1$ experience only a mild
increase in their eccentricity, up to a value $\sim0.2$, while binaries with
smaller $q$ can reach $e>0.8$. Binaries with $e_0>0.3$ tend to reach
eccentricities larger than $0.9$ regardless on $q$, $M_1$ and $\gamma$. 
The evolution 
of $H$ is basically unaffected by the eccentricity of the system in the
bound and unbound phases (in general the average energy subtracted to the 
binary by the star is not affected by the binary eccentricity, see e.g. SHM06), 
while $K$ is interestingly larger for higher $e_0$ when $q$ is large, and
viceversa, it decreases with increasing $e_0$ for small $q$. 

The impact of the assumed slope $\gamma$ of the density profile is highlighted
in figure \ref{figgamma} for binaries with $M_1=10^6\msun$ and $e_0=0.1$. In
general, MBHBs in steeper cusps evolve faster, but to a lower maximum 
eccentricity, during the scattering phase. As explained in SHM08 this is 
because, in the scattering process, stars with $a_*>a$ tend to increase $e$,
while stars with $a_*<a$ tend to decrease it, and the relative weight of
the formers is larger in shallower cusps. Moreover, by increasing $\gamma$, 
the dynamical range covered by the scattering process is much shorter 
(especially for low $q$), because the $a_0-a_{\rm gw}$ gap is smaller, and 
there is less room for significant eccentricity growth. The $K$ rates are 
mildly affected by $\gamma$, with higher value of $\gamma$ resulting 
in smaller $K$, as explained before. Also notice that the absolute value
of $H$ in the bound phase increases a lot with $\gamma$. This is 
because the value of $a_0$ is much smaller for high $\gamma$, and the 
shrink in the bound phase is accordingly much faster. In general, for 
any value of $\gamma$, the eccentricity reached at the end of the 
star scattering phase is $>0.7$ for $q<0.1$, while, again, equal mass
binaries experience a less pronounced increase in the eccentricity.

Figures \ref{figm}, \ref{figecc} and \ref{figgamma} allow also a
detailed study of the evolutionary timescale as a function of the 
system parameters. Since the bound phase is usually much faster
than the unbound one, the evolution timescale of the system is 
set by $a/\dot{a}=\sigma/(G\rho H a)\propto 1/a$. The coalescence
timescale is then set by the bound-GW transition occurring 
at $a_{\rm gw}$. By substituting $a_{\rm gw}$ given by equation (\ref{agw})
in the timescale definition above, we get for the coalescence timescale
\be
\tau_c\propto F(e)^{-1/5}(3-\gamma)^{9/5}q^{-1/5}(1+q)^{2/5}.
\ee
Firstly, we notice that $\tau_c$ is independent on the absolute value of the
MBHB mass, in agreement with figure \ref{figm}. This is a consequence of 
normalizing the stellar distribution outside $r_{\rm inf}$ to an isothermal 
sphere obeying the $M-\sigma$ relation. $\tau_c$ has also a very mild 
dependence on $q$, increasing by a factor of $\sim 3$ when the mass ratio drops
from $q=1$ to $q=1/729$, as shown by the correspondent panels in 
figure \ref{figm}. High values of the maximum eccentricity accelerate
the coalescence by a factor $F(e)^{1/5}$, which is $\sim 5$ for $e=0.9$, 
this effect is clear in the $a$ and $e$ panels of figure \ref{figecc}. 
The impact of $\gamma$ is also quite
mild, in spite of the $9/5$ exponent, and it modifies $\tau_c$ by a factor
of $\sim3-4$ (see $a$ and $e$ panels in figure \ref{figgamma}). 
In general, we find $10^7{\rm yr}<\tau_c<few\times10^8{\rm yr}$. 

\subsection{Comparison with numerical works}
The evolution of MBHBs in stellar environments has been tackled by
several authors by means of full N-body simulations
\citep{mm01,hm02,ar03,mf04,bau06,mat07,mms07,ber09,as09,as10}. However, the 
limited number of particles ($N<10^6$) in such simulations results
in very noisy behaviour for the binary eccentricity, and it is 
difficult to draw conclusions about the general trends behind the 
numerical noise. We can compare our results with N-body simulations
carried out in two regimes: $q=1$ (equal mass inspirals) and 
$q=1/1000$ (intermediate MBH-MBH inspiral). \cite{mm01} 
carried out numerical integration of equal MBHBs embedded in two
merging isothermal cusps ($\gamma=2$). Starting with circular orbits they
find a mild eccentricity increase to a value of $\lsim0.2$ during 
the stellar driven hardening phase, consistent
with our findings. \cite{mms07} considered equal MBHBs embedded in 
Dehnen density profiles \citep{de93} with $\gamma=1.2$ with different 
initial eccentricities.
Again, they find that circular binaries tend to stay circular, while 
eccentric binaries tend to increase their eccentricities in reasonable 
agreement with the prediction of scattering experiments, and, consequently,
with the tracks we presented in figure \ref{figecc} for
the $q=1$ case. Simulations carried by \cite{ar03} and Hamsendorf 
et al. (2002) produce MBHBs with $e_0\approx 0.8$ at the moment of pairing, 
with $e$ subsequently increasing up to $\gsim0.95$, 
again consistent with our findings.
\cite{ber09} studied the evolution of equal MBHBs in rotating systems
described by a King stellar distribution. They also find quite 
eccentric binaries at the moment of pairing ($e>0.4$), and the subsequent
evolution leads to eccentricities larger than 0.95 at which point GW emission
takes over, again consistent with our findings. \cite{as09} focused 
on intermediate MBHBs ($M\sim10^3\msun$) in massive star clusters. They 
employ a machinery similar to ours, coupling full N-body simulations to three 
body scattering experiments.
Their binaries have significant eccentricity ($\sim0.5-0.6$) at the moment 
of pairing, and the predicted range in the {\it LISA} band is $0.1<e<0.3$, 
in good agreement with the results shown by the eccentricity maps in figure
\ref{figlisa}, that we will describe in the next section. 
Similar conclusions are reported by \cite{as10}. On the small $q$ side,
simulations were performed by \cite{bau06} and \cite{mat07}, 
assuming a stellar density profile $\gamma=1.75$. When properly
rescaled, the eccentricity increase found in both papers agrees 
surprisingly well with our predictions based on the hybrid cusp erosion
model. Unfortunately, we did not find  any mention in the N-body literature 
about the eccentricity evolution for intermediate values of $q$ 
($0.01<q<0.1$), and it would be useful to compare our results with 
N-body simulations in this intermediate range. We find, 
however, the overall good agreement, at least in the trends, 
shown at the two extremes of the mass ratio range comforting.    

\section{Eccentricity in the {\it LISA} and PTA windows}

One of the main goals of the present study is to draw sensible predictions
for the eccentricity of MBHBs emitting GWs in the {\it LISA} and in the PTA 
frequency ranges. So far, most of the work related to source modelling,
signal analysis and parameter estimation relied on the assumption of 
circular orbits. This seems reasonable because GW emission
is very efficient in dumpening the binary eccentricity, and
since GW detectors ({\it LISA} in particular) are sensitive to the
very end of the MBHB inspiral, sources are assumed to be circular when they 
enter the observable band. However the level of residual eccentricity
critically depends on how large $e$ is at the transition between 
the stellar hardening and the GW phases. In the scenario proposed here, such  
value can easily be larger than $0.9$, implying non negligible residual 
eccentricities in the frequency bands to be probed by future GW detectors. 

\begin{figure*}
\centering
\begin{tabular}{cc}
\includegraphics[width=0.46\linewidth]{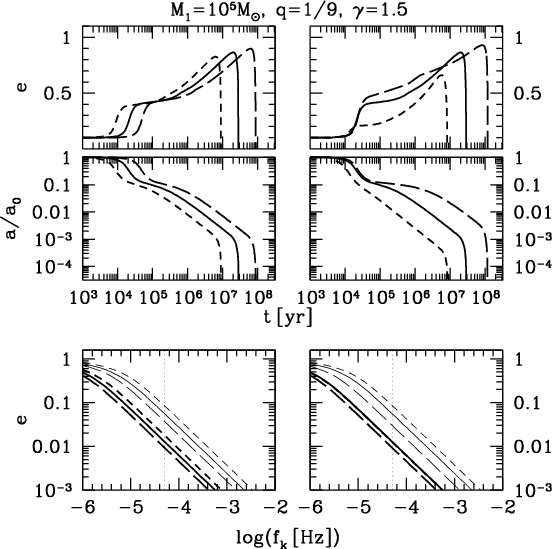} &
\includegraphics[width=0.46\linewidth]{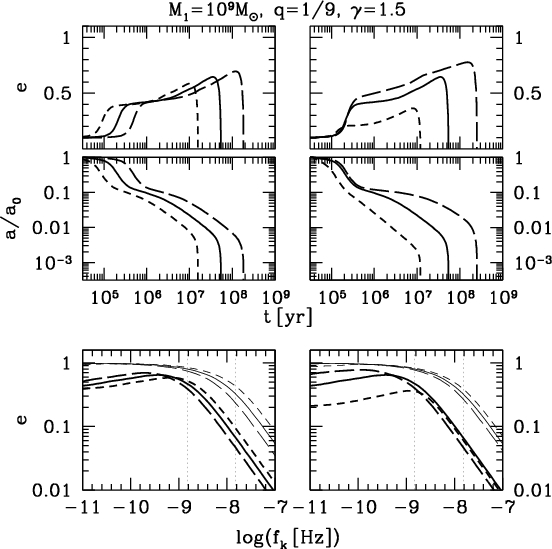} \\
\end{tabular}
\caption{Impact on the $M-\sigma$ assumption and on the $\rho$ normalization 
on the evolution of the binary. {\it Left plot}: representative {\it LISA} 
source, with system parameter highlighted at the top. {\it Right plot}: 
typical PTA sources, with system parameters highlighted at 
the top. In each plot, in the left panels we considered three different
normalizations of the stellar velocity dispersion: $\sigma=\hat{\sigma}$ 
(i.e., the value predicted by the $M-\sigma$ relation, solid lines), 
$1.3\hat{\sigma}$ (short--dashed lines), $0.7\hat{\sigma}$ 
(long--dashed lines); in 
the right panel we stick the efficiency of loss cone refilling to 
$\rho_{\rm inf}$ (solid lines), $10\rho_{\rm inf}$ (short--dashed lines), and
$0.1\rho_{\rm inf}$ (long--dashed lines). Top and middle panels represent the
evolution of $e$ and $a$ against $t$, respectively; bottom panels represent
$e(f_k)$. In these latter panels, $f_{LISA}$ and and the relevant frequency
PTA range $3\times10^{-9}{\rm Hz}<f<3\times10^{-8}{\rm Hz}$ are highlighted.}  
\label{figloss}
\end{figure*}

\subsection{Eccentricity maps}
To convert our 
evolutionary track into predictions for GW observations, we proceed 
as follows. For each MBHB (uniquely defined by $M_1$, $q$, $\gamma$, $e_0$),
we convert the $a/a_0(t)$ tracks into $a(t)$ tracks, and then we 
compute $f_k(t)$, the orbital frequency of the binary, 
simply by assuming Kepler's law. Having $e(t)$ and $f_k(t)$
we then construct the $e(f_k)$ evolution from the moment of pairing to the
final coalescence. We then select two frequencies appropriate for 
{\it LISA} and PTA campaigns, $f_{LISA}$ and $f_{\rm PTA}$ respectively, and 
evaluate $e|_{f_{LISA}}$ and $e|_{f_{\rm PTA}}$. Remember that for circular 
binaries, gravitational radiation is emitted at $f_{\rm gw}=2f_k$. 
We pick $f_{LISA}=5\times 10^{-5}$ Hz (corresponding 
to $f_{\rm gw}=10^{-4}$ Hz, which is approximately the lower bound of the 
{\it LISA} band); on the other hand, we use $f_{\rm PTA}=5\times10^{-9}$ Hz 
(corresponding to $f_{\rm gw}=10^{-8}$ Hz, which is approximately
the frequency at which a 5-to-10 yr PTA campaign will be most sensitive to).
The results are shown in figures \ref{figlisa} and \ref{figpta} as contour
plots  $e|_{f_{LISA}}(M_1,q)$ and $e|_{f_{\rm PTA}}(M_1,q)$, for selected value
of $\gamma$ and $e_0$, as labelled in the figures. Let us start discussing
the {\it LISA} case. Firstly, we limited $M_1$ to an upper value of $10^7\msun$,
since the inspiral of binaries with higher masses will fall outside the
{\it LISA} band {\footnote{The higher frequency signal coming from
the coalescence and ringdown as well as higher harmonic corrections 
to the late inspiral phase \citep{ed08} are likely to push the detectable
mass limit close to $10^8\msun$; however, the imprint of any residual 
eccentricity would be very small and hard to observe for such 
extreme masses.}}. We also excluded from our contour plots systems with
$M_2<100\msun$, for the reasons discussed in Section \ref{caveats}.
The general trend is that lighter unequal mass binaries tend to have
larger $e$ when they enter the relevant frequency range. The mass trend
is easily explained by the fact that our treatment is largely mass
invariant, but the absolute frequency of the system is not! Same 
stages of the MBHB evolution correspond to progressively lower frequencies
as $M_1$ increases, and since we are in the GW dominated phase (and thus 
$de/df<0$), more massive systems have lower eccentricities at a given frequency.
Binaries with smaller $q$ are in general more eccentric because the 
eccentricity growth they experience in the stellar scattering phase is
larger. If binaries are approximately circular at the moment of pairing 
($e_0=0.01$), 
then the maximum eccentricity in the {\it LISA} band is $\sim0.2$ when 
$M_1\sim10^4\msun$ and $q\lsim 0.1$, and the general trend is largely
independent on $\gamma$. This is the result of two competitive effects:
milder cusps lead to larger values of $e$, but in this case GW takes 
over earlier (because the timescale of the 3-body scattering evolution
is set by the density at $r_{\rm inf}$ which is lower for milder central
cusps, being $r_{\rm inf}$ itself larger), and it has more time to
circularize the orbit before the binary get to $f_{LISA}$. 
If binaries are significantly
eccentric at the moment of pairing ($e_0=0.6$ as a study case), then the $q$
dependence in the contour plots almost disappear, because the eccentricity
growth in the scattering phase is very efficient irrespective of $q$ when
$e_0$ is large. In this case, light MBHBs ($M_1<10^4$) may reach $f_{LISA}$
with eccentricities up to $\sim0.5$. 

The situation is even more 'dramatic' 
for PTA observations. PTAs are sensitive to much larger masses ($M_1>10^7\msun$)
emitting at much lower frequencies ($f\approx10^{-8}$ Hz). 
Systems are, in general, caught far from coalescence
(the typical time to coalescence is $\sim10^4$ yr, \citealt{sv10})
and they did not have much time to circularize under the effect of GW emission.
Because of this, even if binaries were circular at the moment of pairing, 
eccentricities can be as high as $0.7$ in the PTA band, with the same trend
observed for the {\it LISA} case (i.e., lighter binaries with smaller $q$ are 
more eccentric). If $e_0=0.6$, then all the systems with $M_1<10^9\msun$
are expected to have $e>0.5$. Again, the results are only mildly dependent 
on $\gamma$. Note that frequencies are
computed in the reference frame of the source, the actual observed 
frequency has than to be appropriately redshifted by a factor $(1+z)$ 
according to the redshift of the emitting system. This means that for
high redshift sources, an {\it observed} frequency of $10^{-4}$ Hz corresponds
to a higher {\it intrinsic} frequency. High redshift sources may then have
milder eccentricities in the observable band, which may be 
relevant for {\it LISA} sources (whereas typical PTA sources are at $z<1$). 

\subsection{The impact of the chosen normalization}

We want to check at this point how the 
$M-\sigma$ normalization and the tuning of the loss cone refilling 
efficiency to $r_{\rm inf}$ impact on our results. Selected cases
of alternative models are presented in figure \ref{figloss}.
The left hand  plot is representative of {\it LISA} sources. 
We see that changing $\sigma$ merely shifts the 
timescale of the binary evolution. For larger $\sigma$, the system
is more compact, the timescale for 3-body scattering is shorter,
GW emission takes over later, and consequently the residual 
$e$ at  $f_{LISA}$ is larger. However, as shown by the $e(f_k)$
tracks, this is at most a factor of 2 effect. Changing the
normalization $\rho$ in the loss-cone refilling process, has instead
a major impact on the evolutionary timescale and on the eccentricity
evolution of the system, but still the residual eccentricity at   
$f_{LISA}$ is basically unaffected when $e_0=0.01$, and it changes
by at most a factor of $3$ in the $e_0=0.6$ case. In the right hand
plot we consider instead the typical PTA source. All the considerations
made for the {\it LISA} case still hold, and in the relevant frequency
range $3\times10^{-9}{\rm Hz}<f<3\times10^{-8}{\rm Hz}$, the expected
$e$ changes by at most a factor of 2. We therefore consider our results 
quite robust irrespective of the assumed normalizations.

\subsection{Eccentricity distributions for selected MBHB population models}

\begin{figure}
\centering
\begin{tabular}{c}
\includegraphics[width=0.85\linewidth]{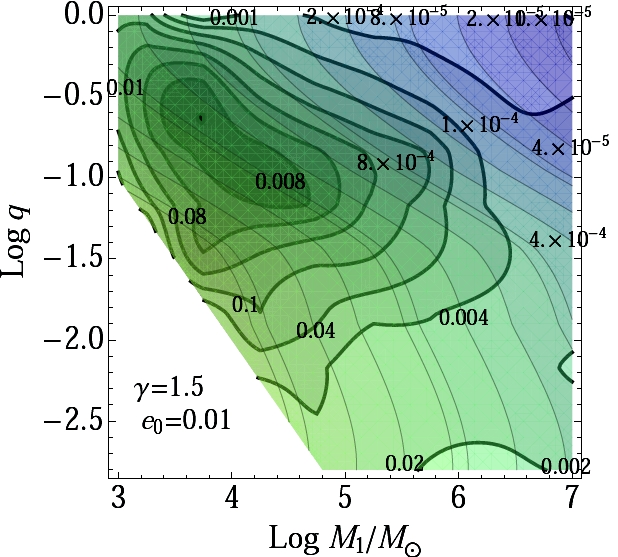}\\
\includegraphics[width=0.85\linewidth]{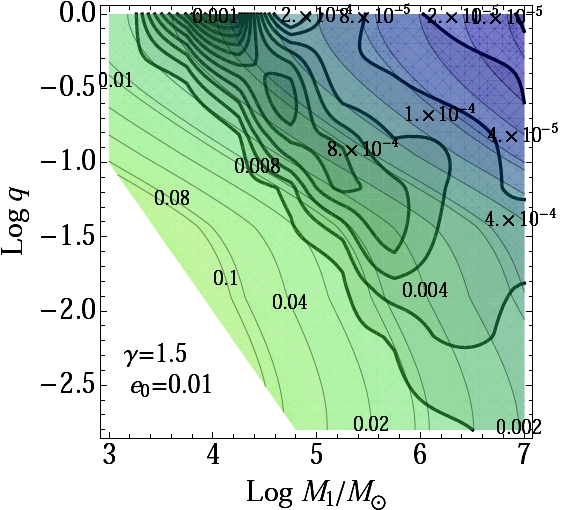}\\
\includegraphics[width=0.85\linewidth]{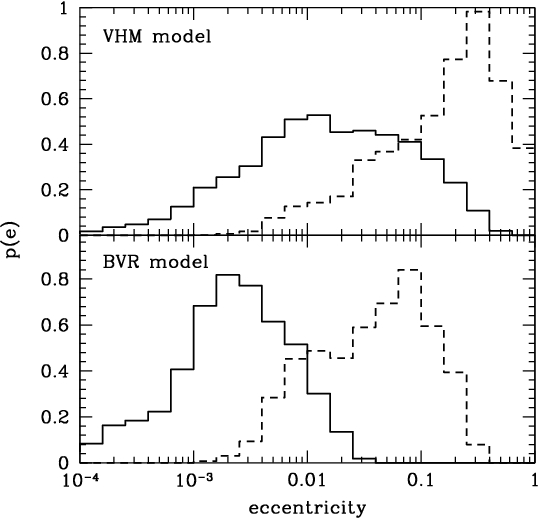}
\end{tabular}
\caption{Evaluation of the eccentricity distribution of MBHBs observed by
{\it LISA}. {\it Top and middle panels}, contour plots of the differential
distribution of observable sources $d^2N/dM_1dq$ (gray scale, contour 
normalization unnecessary for illustrative purposes), superimposed
to the contour plots of $e|_{f_{LISA}}$ (color scale; $\gamma=1.5$, $e_0=0.01$) 
in the $(M_1,q)$ plane.
Inset labels refer to the $e$ contours. Top panel is for the VHM model, middle
panel is for the BVR model. {\it Bottom panel}: probability density function
$p(e)$ corresponding to the contour plot convolution; solid histograms are for
$e_0=0.01$ and dashed histograms are for $e_0=0.6$.}
\label{fecclisa}
\end{figure}
\begin{figure}
\centering
\begin{tabular}{c}
\includegraphics[width=0.85\linewidth]{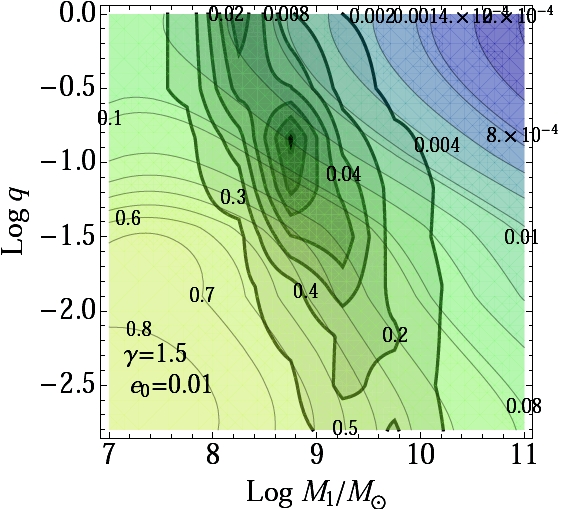}\\
\includegraphics[width=0.85\linewidth]{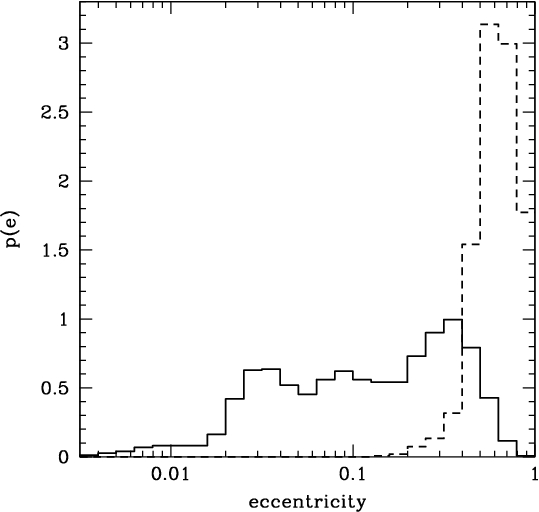}
\end{tabular}
\caption{Same as figure \ref{fecclisa}, but for PTA observations. In the top
panel, the differential distribution of observable sources $d^2N/dM_1dq$ is 
now superposed to $e|_{f_{\rm PTA}}$, again assuming $\gamma=1.5$ and $e_0=0.01$.
The resulting $p(e)$ is plotted in the bottom panel assuming $e_0=0.01$ (solid
histogram) and $e_0=0.6$ (dashed histogram)}
\label{feccpta}
\end{figure}
As a final step, we quantify the eccentricity distribution of GW sources 
in the relevant frequency range resulting by applying our eccentricity 
evolution scheme to standard MBHB population models. 

For the {\it LISA}
case we use two of the models utilized by the {\it LISA} parameter estimation
task force \citep{aru09}: in the first case
seeds are light \citep[$M\gsim100\msun$, VHM model;][]{vhm03}, 
being the remnant of the first POPIII star explosions \citep{mr01}; 
in the second case, already quite heavy ($M\gsim10^4\msun$) seed BHs form by
direct collapse of massive protogalactic discs \citep[BVR model;][]{bvr06}. 
We ran 50 Monte Carlo realizations of each model,
producing 50 catalogues of coalescing binaries over a period of 3 yrs. 
We then estimate the signal-to-noise ratio (SNR) of each binary 
in the {\it LISA} detector by assuming circular inspiral and computing 
the waveform to the 2PN order. We then consider only those events 
resulting in an SNR$>8$ in the detector, and we compute the expected 
eccentricity distribution at  $f_{LISA}$. Results are shown in figure
\ref{fecclisa}. In the upper and in the middle panels we plot the contour
plots of the differential distribution of GW sources as a 
function of $M_1$ and $q$, $d^2N/dM_1dq$, averaged over the 50 Monte Carlo
realizations, superposed to the contour plots for $e|_{f_{LISA}}(M_1,q)$.
The two observed MBHB populations are extremely different:
in the VHM model, the bulk of sources have $M\sim10^4\msun$ with $q\sim0.1$;
while in the BVR model, most of the sources have $M>10^4\msun$ and 
$q\approx 1$. The resulting eccentricities distributions are plotted 
in the lower panel, where we plot the probability density function
$p(e)$ against $e$ for the observed population. When $e_0=0.01$
(binaries approximately circular), eccentricity
is expected to be $<10^{-2}$ in the BVR model, with a peak at about 
$2\times 10^{-3}$, but a broad eccentricity spectrum covering the range 
$10^{-3}-0.2$ is expected in the VHM case. In this latter scenario, infact,
sources are on average less massive and with low $q$, a condition that
maximizes the eccentricity increase during the stellar scattering phase.
If $e_0$ is already large (0.6 in our study case), then the observed
eccentricity at $f_{LISA}$ is peaked at $\sim0.1$ for the BVR case and at
$\sim0.4$ in the VHM case. 

In exploring the consequences for PTA observations, we adopt the standard
Tu-SA population model employed by \cite{svv09}, where merging 
galaxies are populated by MBHs according to the $M-M_{\rm bulge}$ in 
the form given by \cite{tundo07} and accretion is triggered onto
the more massive black hole {\it before} the final coalescence. The 
reader is referred to \cite{svv09} for details. We ran 50 Monte Carlo 
realizations of the model (assuming binaries in circular orbit)  and we pick 
only the individually resolvable sources generating a timing residual larger
than 1ns.  Again, the obtained differential distribution of the individually
resolvable sources, $d^2N/dM_1dq$, averaged over the 50 Monte Carlo
realizations, is superposed to the contour plots for $e|_{\rm PTA}(M_1,q)$
in figure \ref{feccpta}. The source distribution is strongly peaked
around $M_1=10^9\msun$ and $q=0.1$, with a long tail extending to $q=10^{-3}$.
If binaries are approximately circular at the moment of pairing 
($e_0=0.01$), then the expected $p(e)$ is basically flat in the range 
$[0.03, 0.3]$, while for $e_0=0.6$, $p(e)$ has a sharp peak in the range
$[0.5,0.7]$, highlighting the possible significant impact of eccentricity 
for PTA observations.

\section{Discussion and conclusions}
We studied the semimajor axis and eccentricity evolution of massive black 
hole binaries in stellar environments by coupling the results of numerical
3-body scattering experiments to an analytical framework describing the 
evolution of the stellar distribution and the supply of stars to the 
binary loss cone. Our treatment takes into account the scattering of 
bound stars determining the erosion of the stellar cusp bound to the binary,
and the subsequent scattering of unbound stars fed to the binary
loss cone by relaxation processes. We {\it do not} address the nature 
of the relaxation processes leading to loss cone replenishment, but 
we treat the loss cone refilling efficiency as a parameter of the model.
Eventually, GW emission takes over, leading to the final coalescence of
the system. 

Our main finding is that 3-body scattering induces a significant increase
in the MBHB eccentricity, that is not efficiently washed out by GW-induced 
circularization before the system enter the {\it LISA} or the PTA bands.
The eccentricity growth is in general larger for binaries with smaller
mass ratios, and at the stellar scattering-GW transition can easily be
higher than $0.9$. Equal mass binaries in general experience a milder 
eccentricity growth when the initial eccentricity is close to zero.
The eccentricity growth is more prominent for systems characterized by 
smaller masses. Binaries with significant initial eccentricity $e_0>0.3$ 
end up in very eccentric orbits ($e>0.9$) regardless on the other system 
parameters. The impact of the cusp slope can be significant, with shallower
cusps leading to higher maximal values of $e$, as explained in SHM08.
In general, the eccentricity growth is dominated by the bound scattering
phase for binaries with $q<0.1$ and by the unbound scattering phase for 
binaries with larger mass ratios. When compared to the sparse results 
of full N-body simulations found in literature, the results of our models
are in reasonable agreement with those of numerical studies. 

The implications for GW observations are relevant. When binaries are 
circular at the moment of pairing, their eccentricity when they enter
the {\it LISA} band is in the range $10^{-5}-0.2$, and it is larger for 
low mass unequal binaries. If binaries are already eccentric at the moment
of pairing, these figures shift to the range $10^{-3}-0.5$, with lower 
mass binaries leading to higher eccentricities and only a mild dependence
on the mass ratio. We emphasize once again that in our treatment, 
the total mass of the system sets the the typical scale of the problem;
because of this, the residual eccentricity in the {\it LISA} band
is {\it larger for lighter binaries}. This is important because {\it LISA} 
will be mostly sensitive to low mass MBHBs in the range $10^4-10^5\msun$.
In the PTA windows, the implications are even stronger.  
Initially circular systems end up with eccentricities in the range 
$10^{-3}-0.8$ at a frequency of $10^{-8}$ Hz (relevant to PTA observations),
and for significant initial eccentricity, binaries with $M_1<10^{-9}\msun$ 
always have $e>0.5$ in the PTA band. The trend with the mass and the
mass ratio are the same as for their {\it LISA} counterparts. All the 
results are basically independent on the cusp slope $\gamma$, and are
only mildly dependent on the 
normalization of the stellar density distribution, and on the efficiency
of the loss cone refilling. 

Once applied to standard MBHB population models, these results predict
eccentricities in the range $10^{-3}-0.2$ (depending on the adopted 
seed formation model) for observable {\it LISA} sources, and a broad 
flat $e$ distribution in the interval $0.03-0.3$ for source individually
resolvable by  PTAs. High initial values of $e$, naturally lead to more
eccentric systems.

Our results are of particular interest for the GW community, showing
that a proper treatment of the eccentricity might be crucial in the 
challenge of GW detection. Mock data challenge initiatives like the
{\it LISA} mock data challenge, have so far implemented circular MBHBs
only, and consequently, the ability of data analysis
and parameter estimation algorithms has been proven only in this 
situation. The typical eccentricity values found in the {\it LISA}
band ($<0.2$ for systems in circular orbit at the moment of pairing)
allow for a perturbative approach to the problem of constructing 
trustwhorty post-Newtonian waveform, as the one recently employed 
by \cite{nico09}. In the light of the results presented here, 
further work in this direction would be extremely valuable. 
The addition of a non zero eccentricity would affect the
waveform by adding significant amplitude modulation and phase 
precession, which in turn would affect our detection and parameter
estimation ability (work in this direction is ongoing and
preliminary results can be found in \cite{ps10}).
Also in the PTA source modelling field, the assumption of circular orbits
has been widely used so far, with the notable exception of \cite{en07}. 
Further work on eccentric source modelling is needed, 
in order to address properly how an eccentric population of MBHB would
affect the overall level of the background, the statistics of individually 
resolvable sources, the detailed shape of the residuals and our ability 
of extracting signals and estimating source parameters.

We finally stress that our model is oversimplified, relying only on 
stellar dynamics without taking into account the possible impact 
of the presence of large amounts of gas surrounding the binary. 
Gas dynamics may be particularly relevant to {\it LISA} sources, which
are expected to be found in mergers of small galaxies at high redshift \citep
{svh07}, where the mass content of galaxies is likely to be dominated by gas
\citep[see, e.g.,][]{c00}. In 
this view, our model should provide a more trustful description
of PTA sources which consist instead of massive binaries at low redshift
\citep{svv09}, 
likely hosted by gas poorer galaxies. Nonetheless, we should bear in mind
that recent studies also found significant eccentricity increase 
in MBHBs driven by circumbinary disks \citep{an05,cua09}. Moreover 
gas dynamics may not be efficient enough to drive the final coalescence
of MBHB systems \citep{lod09}, and also for low mass sources at high redshift,
stellar dynamics may provide a viable alternative path through the final
coalescence. We hope that our exploratory study will stimulate further 
research on the subject, which is of critical importance for a comprehensive
modelling of MBHB evolution and for their future observation in the 
upcoming gravitational radiation windows.  

\acknowledgments
I am grateful to E. Berti, M. Dotti and E. Porter for their comments and 
suggestions, and to Frank Ohme for the invaluable help in constructing the
contour plots shown in the paper. 



\begin{thebibliography}{}

\bibitem[Aarseth (2003)]{ar03} Aarseth S. J., 2003, Ap\&SS, 285, 367

\bibitem[Armitage \& Natarajan (2005)]{an05} Armitage P. J. \& Natarajan P., 2005, Apj, 634, 921

\bibitem[Arun et. al (2009)]{aru09} Arun K. G. et al., 2009, CQGra, 26, 4027

\bibitem[Amaro-Seoane, Miller \& Freitag (2009)]{as09} Amaro-Seoane P., Miller M. C. \& Freitag M., 2009, ApJ, 692,50

\bibitem[Amaro-Seoane et al. (2010)]{as10} Amaro-Seoane P., Eichhorn C., Porter E. K. \& Spurzem R., 2010, MNRAS, 401, 2268

\bibitem[Babak et al. (2009)]{bab09} Babak S. et al., 2009, arXiv:0912.0548

\bibitem[Baumgardt, Gualandris \& Portegies Zwart (2006)]{bau06} Baumgardt H., Gualandris A. \& Portegies Zwart S., 2006, MNRAS, 372, 174 

\bibitem[Begelman, Blandford \& Rees (1980)]{br80} Begelman M. C., Blandford R. D. \& Rees M. J., 1980, Nature, 287, 307

\bibitem[Begelman, Volonteri \& Rees (2006)]{bvr06} Begelman M. C., Volonteri \& Rees M. J., 2006, MNRAS, 370, 289

\bibitem[Berczik et al. (2006)]{ber06} Berczik P., Merritt D., Spurzem R. \& Bischof H. P., 2006, ApJ, 642, 21

\bibitem[Berentzen et al. (2009)]{ber09} Berentzen I., Preto M., Berczik P., Merritt D. \& Spurzem R., 2009, ApJ, 695, 455

\bibitem[Binney \& Tremaine (1987)]{bt87} Binney J. \& Tremaine S., "Galactig Dynamics", Princeton University Press  

\bibitem[Chen et al. (2009)]{chen09} Chen X., Madau P., Sesana A. \& Liu F. K., 2009, ApJ, 697, 149

\bibitem[Chen et al. (2010)]{chen10} Chen X., Sesana A., Madau P. \& Liu F. K., 2010, submitted to ApJ.

\bibitem[Cohn \& Kulsrud (1978)]{ck78} Cohn H. \& Kulsrud R. M., 1978, ApJ, 226, 1087

\bibitem[Cole et al. (2000)]{c00} Cole S., Lacey C. G., Baugh C. M. \& Frenk C. S., 2000, MNRAS, 319, 168

\bibitem[Colpi, Mayer \& Governato (1999)]{cmg99} Colpi M., Mayer L. \& Governato F., 1999, ApJ, 525, 720

\bibitem[Cuadra et al. (2009)]{cua09} Cuadra J., Armitage P. J., Alexander R. D. \&  Begelman M. C., 2009, MNRAS, 393, 1423

\bibitem[Danzmann et al. (1998)]{dan08} Danzmann K. et al., 1998, LISA-Laser Interferometer Space Antenna, Pre-Phase â A Report, 2nd edn. Max-Planck-Institute fur Quantenaptik, Garching

\bibitem[Dehnen (1993)]{de93} Dehnen W., 1993, MNRAS, 265, 250

\bibitem[Dehnen \& Binney (1998)]{db98} Dehnen W. \& Binney J., 1998, MNRAS, 298, 387

\bibitem[Dotti, Colpi \& Haardt (2006)]{dotti06} Dotti M., Colpi M. \& Haardt F., 2006, MNRAS, 367, 103

\bibitem[Dotti et al. (2007)]{dotti07} Dotti M., Colpi M., Haardt F. \&  Mayer L., 2007, MNRAS, 379, 956

\bibitem[Enoki et al. (2004)]{eno04} Enoki M., Inoue K. T., Nagashima M. \& Sugiyama N., 2004, ApJ, 615, 19

\bibitem[Enoki \& Nagashima (2007)]{en07} Enoki M. \& Nagashima M., 2007, PThPh, 117, 241

\bibitem[Escala et al. (2005)]{escala05} Escala A., Larson R. B., Coppi P. S. \&  Mardones D., 2005, ApJ, 630, 152

\bibitem[Ferrarese et al. (1994)]{f94} Ferrarese L., van den Bosch F. C., Ford H. C., Jaffe W. \& O'Connell R. W., 1994, AJ, 108, 2403	

\bibitem[Frank \& Rees (1976)]{fr76} Frank J. \& Rees M. J., 1976, MNRAS, 176, 633

\bibitem[Gebhardt et al. (1996)]{g96} Gebhardt K. et al., 1996, AJ, 112, 105

\bibitem[Haring \& Rix (2004)]{hr04} Haring N. \& Rix, H. W., 2004, ApJ, 604, 89

\bibitem[Haehnelt \& Rees (1993)]{hr93} Haehnelt M. G. \& Rees M. J., 1993, MNRAS, 263, 168

\bibitem[Haehnelt (1994)]{Haehnelt94} Haehnelt M. G., 1994, MNRAS, 269, 199

\bibitem[Hemsendorf, Sigurdsson \& Spurzem (2002)]{hm02} Hemsendorf M., Sigurdsson S. \& Spurzem R., 2002, ApJ, 581, 1256

\bibitem[Jaffe \& Backer (2003)]{Jaffe03} Jaffe A. H. \&  Backer D. C., 2003, ApJ, 583, 616

\bibitem[Janssen et al. (2008)]{Janssen08} Janssen G. H., Stappers B. W., Kramer M., Purver M., Jessner A. \&  Cognard I., 2008, in ``40 YEARS OF PULSARS: Millisecond Pulsars, Magnetars and More'', AIP Conference Proceedings, 983, 633

\bibitem[Jenet et al. (2005)]{Jenet05} Jenet F. A., Hobbs G. B., Lee K. J. \&  Manchester R. N., 2005, ApJ, 625, 123

\bibitem[Jenet et al. (2009)]{jen09} Jenet R. et al., 2009, arXiv:0909.1058

\bibitem[Lauer et al. (1995)]{l95} Lauer T. R. et al.,  1995, AJ, 110, 2622 

\bibitem[Lazio (2009)]{Lazio09} Lazio J., 2009, arXiv:0910.0632

\bibitem[Lightman \& Shapiro (1977)]{ls77} Lightman A. P. \& Shapiro S. L., 1977, ApJ, 211, 244

\bibitem[Lodato et al. (2009)]{lod09} Lodato G., Nayakshin S., King A. R. \& Pringle J. E., 2009, MNRAS, 398, 1392

\bibitem[Madau \& Rees (2001)]{mr01} Madau P. \& Rees M. J., 2001, ApJ, 551, 27

\bibitem[Magorrian et al. (1998)]{mago98} Magorrian J. et al., 1998, AJ, 115, 2285

\bibitem[Makino \& Funato (2004)]{mf04} Makino J. \& Funato Y., 2004, ApJ, 602, 93

\bibitem[Manchester (2008)]{Manchester08} Manchester R. N., 2008, in ``40 YEARS OF PULSARS: Millisecond Pulsars, Magnetars and More'', AIP Conference Proceedings, 983, 584

\bibitem[Matsubayashi, Makino \& Ebisuzaki (2007)]{mat07} Matsubayashi T., Makino J. \& Ebisuzaki T., 2007, ApJ, 656, 879

\bibitem[Merritt, Mikkola \& Szell (2007)]{mms07} Merritt D., Mikkola S. \& Szell A., 2007, ApJ

\bibitem[Merritt \& Poon (2004)]{mp04} Merritt D. \& Poon M. Y., 2004, ApJ, 606, 788

\bibitem[Merritt \& Quinlan (1998)]{mq98} Merritt D. \& Quinlan G. D., 1998, NewA, 498, 625 

\bibitem[Merritt \& Szell (2006)]{ms06} Merritt D. \& Szell A., 2006, ApJ, 648, 890

\bibitem[Milosavljevic \& Merritt (2001)]{mm01} Milosavljevic M. \& Merritt D., 2001, ApJ, 563, 34

\bibitem[Milosavljevic \& Merritt (2003)]{mm03} Milosavljevic M. \& Merritt D., 2001, ApJ, 596, 860

\bibitem[Mikkola \& Valtonen (1992)]{mv92} Mikkola S. \& Valtonen M.J., 1992, MNRAS, 259, 115

\bibitem[Perets \& Alexander (2008)]{pa08} Perets H. B. \& Alexander T., 2008, ApJ, 677, 146

\bibitem[Peters \& Mathews (1963)]{pm63} Peters P. C. \& Mathews J., 1963, PhRv, 131, 435

\bibitem[Porter \& Cornish (2008)]{ed08} Porter E. K. \& Cornish N. J., PhRvD, 78, 4005

\bibitem[Porter \& Sesana (2010)]{ps10} Porter E. K. \& Sesana A., submitted to PhRvD, arXiv:1005.5296

\bibitem[Quinlan (1996)]{qui96} Quinlan G. D., 1996, NewA, 1, 35 

\bibitem[Rest et al. (2001)]{r01} Rest A., van den Bosch F. C., Jaffe W., Tran H., Tsvetanov Z., Ford H. C., Davies J. \& Schafer J., 2001, AJ, 121, 2431

\bibitem[Rhook \& Wyithe (2005)]{rw05} Rhook K. J. \& Wyithe J. S. B., 2005, MNRAS, 361, 1145 

\bibitem[Sesana et al. (2004)]{Sesana04} Sesana A., Haardt F., Madau P. \&  Volonteri M., 2004, ApJ, 611, 623

\bibitem[Sesana et al. (2005)]{Sesana05} Sesana A., Haardt F., Madau P. \&  Volonteri M., 2005, ApJ, 623, 23

\bibitem[Sesana, Haardt \& Madau (2006)]{shm06} Sesana A., Haardt F. \& Madau P., 2006, ApJ, 651, 392, SHM06

\bibitem[Sesana, Haardt \& Madau (2008)]{shm08} Sesana A., Haardt F. \& Madau P., 2008, ApJ, 686, 432, SHM08

\bibitem[Sesana, Vecchio \& Colacino (2008)]{svc08} Sesana A., Vecchio A. \& Colacino C. N., 2008, MNRAS, 390, 192

\bibitem[Sesana, Vecchio \& Volonteri (2009)]{svv09} Sesana A., Vecchio A. \&  Volonteri M., 2009, MNRAS, 384, 2255 

\bibitem[Sesana \& Vecchio (2010)]{sv10} Sesana A. \& Vecchio A., 2010, submitted to Phys. Rev. D, arXiv:1003.0677

\bibitem[Sesana, Volonteri \& Haardt (2007)]{svh07} Sesana A., Volonteri M. \&  Haardt F., 2007, MNRAS, 377, 1711 

\bibitem[Tremaine et al. (2002)]{tr02} Tremaine, S., et al., 2002, ApJ, 574, 740

\bibitem[Tundo et al. (2007)]{tundo07} Tundo E., Bernardi M., Hyde J. B., Sheth R. K. \& Pizzella A., ApJ, 663, 53

\bibitem[Volonteri, Haardt \& Madau (2003)]{vhm03} Volonteri M., Haardt F. \& Madau P., 2003, ApJ, 582, 599 

\bibitem[White \& Rees (1978)]{wr78} White S. D. M. \& Rees M. J., 1978, MNRAS, 310, 645

\bibitem[Wyithe \& Loeb (2003)]{Wyithe03} Wyithe J.~S.~B. \& Loeb A., 2003, ApJ, 590, 691

\bibitem[Yu (2002)]{yu02} Yu Q., 2002, MNRAS, 331, 935

\bibitem[Yunes et al. (2009)]{nico09} Yunes N,, Arun K. G., Berti E. \& Will C. M., 2009, PhRvD, 80, 4001

\end{thebibliography}
\end{document}